\documentclass{jaa}
\usepackage{natbib}
\usepackage{amssymb}
\usepackage{url} 
\usepackage{aas_macros}
\usepackage{enumitem}
\setlist{nolistsep}
\usepackage{graphicx}
\usepackage[hidelinks]{hyperref}
\usepackage{xcolor}
\hypersetup{
    colorlinks,
    linkcolor={red!50!black},
    citecolor={blue!50!black},
    urlcolor={blue!80!black}
    }
\usepackage{subfiles}

\begin{document}\sloppy
\title{Development of a Compton Imager Setup}


\author{Anuraag Arya\textsuperscript{1,*}, Harmanjeet Singh Bilkhu\textsuperscript{1}, Sandeep Vishwakarma\textsuperscript{3}, Hrishikesh Belatikar\textsuperscript{1}, Varun Bhalerao\textsuperscript{1}, Abhijeet Ghodgaonkar\textsuperscript{4}, Jayprakash G. Koyande\textsuperscript{3}, Aditi Marathe\textsuperscript{1}, N. P. S. Mithun\textsuperscript{5}, Sanjoli Narang\textsuperscript{2,6}, Sudhanshu Nimbalkar\textsuperscript{2}, Pranav Page\textsuperscript{2}, Sourav Palit\textsuperscript{7}, Arpit Patel\textsuperscript{5}, Amit Shetye\textsuperscript{2} Siddharth Tallur\textsuperscript{2}, Shriharsh Tendulkar \textsuperscript{3}, Santosh Vadawale\textsuperscript{5} and Gaurav Waratkar\textsuperscript{1}}

\affilOne{\textsuperscript{1}Department of Physics, IIT Bombay, Powai, Mumbai -- 400076, India.\\}
\affilTwo{\textsuperscript{2}Department of Electrical Engineering IIT Bombay, Powai, Mumbai -- 400076, India.\\}
\affilThree{\textsuperscript{3}Department  of Astronomy and Astrophysics, TIFR, Colaba, Mumbai -- 400005, India.\\}
\affilFour{\textsuperscript{4}Department  of High Energy Physics, TIFR, Colaba, Mumbai -- 400005, India.\\}
\affilFive{\textsuperscript{5}Astronomy and Astrophysics division, PRL, Ahmedabad -- 380054, India.\\}
\affilSix{\textsuperscript{6}Computer Science \& Artificial Intelligence Laboratory, MIT, USA.\\}
\affilSeven{\textsuperscript{7}ICSP, Netai Nagar, Kolkata -- 700099, India.}

\newcommand{\asat}{\emph{AstroSat}}
\newcommand{\stcb}{detector board}
\newcommand{\element}[2]{\ensuremath{^{#1}\texttt{#2}}}
\newcommand{\ee}[1]{\ensuremath{\times 10^{#1}}}
\newcommand{\cps}{\ensuremath{\mathrm{counts~s}^{-1}}}
\newcommand{\degr}{\ensuremath{^\circ}}

\newcommand{\vbdone}{}
\newcommand{\update}[1]{{#1}}
\newcommand{\updatetwo}[1]{{ #1}}
\newcommand{\updatethird}[1]{ {#1} }

\newcommand\scalemath[2]{\scalebox{#1}{\mbox{\ensuremath{\displaystyle #2}}}}


\twocolumn[{

\maketitle

\corres{arya.a@iitb.ac.in}


\begin{abstract}
Hard X-ray photons with energies in the range of hundreds of keV typically undergo Compton scattering when they are incident on a detector. In this process, an incident photon deposits a fraction of its energy at the point of incidence and continues onwards with a change in direction that depends on the amount of energy deposited. By using a pair of detectors to detect the point of incidence and the direction of the scattered photon, we can calculate the scattering direction and angle. The position of a source in the sky can be reconstructed using many Compton photon pairs from a source. We demonstrate this principle in the laboratory by using a pair of Cadmium Zinc Telluride (CZT) detectors sensitive in the energy range of 20--200~keV, \update{similar to those used in \asat/CZT Imager (CZTI)}. The laboratory setup consists of the two detectors placed perpendicular to each other in a lead-lined box. The detectors are read out by a custom-programmed \update{Xilinx PYNQ-Z2 FPGA board}, and data are then transferred to a \update{personal computer (PC)}. \update{There are two key updates from CZTI: the detectors are read concurrently rather than serially, and the time resolution has been improved from $20~\mu$s to $7.5~\mu$s.}
We irradiated the detectors with a collimated \element{133}{Ba} source and identified Compton scattering events for the 356~keV line. We run a Compton reconstruction algorithm to correctly infer the location of the source in the detector frame, \update{with a location-dependent angular response measure of 16\degr--30\degr}. This comprises a successful technology demonstration for a Compton imaging camera in the \update{hard X-ray} regime. We present the details of our setup, the data acquisition process, and software algorithms, and showcase our results. \update{We also quantify the limitations of this setup and discuss ways of improving the performance in future experiments.}
\end{abstract}

\keywords{X-ray imaging---hard X-rays---CZT detectors---X-ray telescopes.}

}]

\doinum{12.3456/s78910-011-012-3}
\artcitid{\#\#\#\#}
\volnum{000}
\year{0000}
\pgrange{1--}
\setcounter{page}{1}
\lp{15}

\section{Introduction} \label{sec:intro}

High Energy photons in the energy range from hundreds of keV to tens of MeV typically undergo Compton \updatethird{scattering} in a detector --- depositing a fraction of their energy at the interaction site, and creating a secondary photon, which may in turn be detected by the same or another detector \citep{Compton1923,knoll}. The scattered photon has a lower energy, and typically deposits its full energy in the detector via photoelectric interactions. The locations of the first and second interactions, and the energy deposits at both places, can be used to constrain the direction of the photon source. This concept is utilized for ``Compton Imaging'' \citep{1989A&A...221..396V,2000A&AS..145..311B}. This technique forms the core of astrophysical missions starting with COMPTEL~\citep{schonfelder1993} to the upcoming Compton Spectrometer and Imager~\citep[COSI;][]{tomsick2019cosi}. However, past missions have typically focused on the Gamma-ray band, and operated in the MeV range \citep{kieransthesis,kierans2022}.

\update{There is growing interest in extending Compton imaging to the hard X-ray regime (hundreds of keV), where a significant fraction of photons still undergo Compton scattering before being fully absorbed via photoelectric interactions \citep{knoll}. For instance, COSI is sensitive to photons down to 200~keV, to ensure coverage of the sub-MeV region \citep{2024icrc.confE.745T}. However, COSI and similar missions typically use Germanium detectors, which need to be operated at cryogenic conditions and thereby increase instrument complexity. On the other hand, the lower end of this energy range can conveniently be studied with Cadmium Zinc Telluride (CZT) detectors, which can be operated at room temperature \citep[see for instance][]{nandi2009indianpayloadsrt2experiment}
.}

\update{\citet{yabu2021tomographic} have demonstrated the operation of such a sub-MeV Compton camera using Cadmium Telluride (CdTe) and Silicon detectors, with a focus on near-field three-dimensional reconstruction of the source. In this work, we discuss our first steps towards building such an instrument for astrophysical applications, by developing and testing a laboratory model of a two-detector CZT-based Compton Imager. We irradiate the setup with radioactive sources and can successfully reconstruct the source position using observed data.}

This paper is organised as follows. In \S\ref{sec:setup} we describe our experimental setup. \S\ref{sec:data} deals with detector calibration and data acquisition. We present our analysis and source position reconstruction in \S\ref{sec:reconstruct}, and conclude with a summary and future prospects in \S\ref{sec:conc}.

\section{Experimental setup}\label{sec:setup}
Our experimental setup consists of three main components: the detectors (\S\ref{subsec:det}), power and readout electronics (\S\ref{subsec:elec}), and the enclosure (\S\ref{subsec:mech}).

\subsection{Detectors}\label{subsec:det}

\begin{figure}[!thp]
    \centering
        \centering
        \includegraphics[width=0.9\columnwidth]{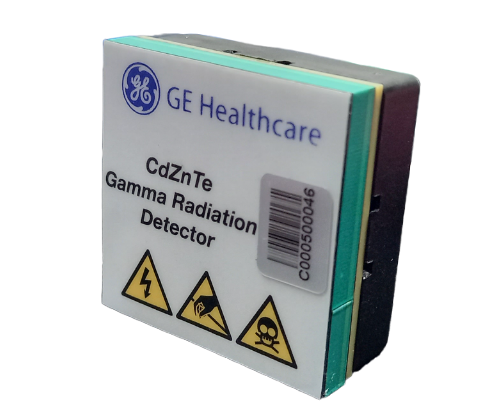}
        \caption{A Cadmium Zinc Telluride (CZT) detector used in our experiments.  These detectors are manufactured by GE Medical (formerly Orbotech).}
        \label{fig:det_image}
\end{figure}

 Our setup is based on 5~mm thick CZT detectors (Figure~\ref{fig:det_image}) manufactured by GE Healthcare\footnote{\url{https://www.gehealthcare.com/products/molecular-imaging/nuclear-medicine/nm-ct-870-czt}.} (formerly Orbotech Medical Solutions). The detectors consist of a 16$\times$16 grid of square pixels with a total size of 39~mm on a side. They operate in a nominal energy range of 20--200~keV, with an energy resolution of 10--12\% at 60~keV \citep{2011ExA....29...27K,Bhalerao2017}. \update{The average detector composition is Cd$_{0.9}$Zn$_{0.1}$Te$_{1.0}$, giving a band gap of 1.57~eV \citep{s24030725}, and an average electron-hole pair creation energy of 4.43~eV\footnote{More details are available at \url{https://astrosat.iucaa.in/czti/sites/default/files/documents/cztdocs/CZT-GEOMETRIC_DETAILS.pdf}.}.}

\update{Such detectors have been successfully used in various Indian space-based hard X-ray instruments before, including RT-2  \citep{nandi2009indianpayloadsrt2experiment}, the Cadmium Zinc Telluride Imager on board \asat\ \citep[\asat/CZTI;][]{Bhalerao2017}, and the High Energy L1 Orbiting X-ray Spectrometer on \emph{Aditya-L1} \citep[HEL1OS;][]{2017CSci..113..625S}.}

The connections to these detectors are low voltage power, a high voltage negative bias, and communication lines. \update{We operate these detectors at $-600$~V, following \asat/CZTI.} The detector module includes circuitry to 
detect on-board triggers, amplify them with a charge-sensitive pre-amplifier, and digitize the output. The digital output is communicated via Low Voltage Differential Signaling (LVDS) format on Serial Peripheral Interface (SPI) lines.


\subsection{Electronics}\label{subsec:elec}
\begin{figure*}[htb]
    \centering
    \includegraphics[width=2\columnwidth]{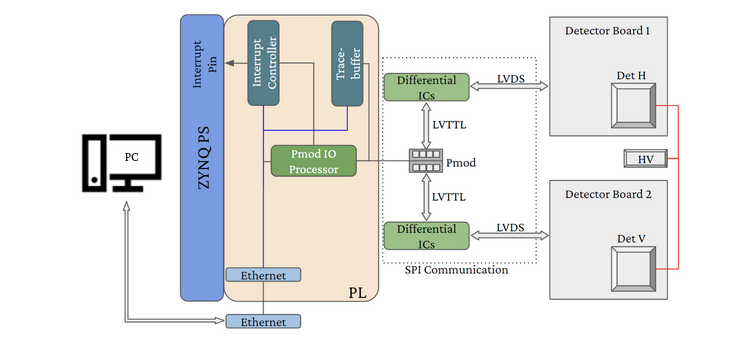}
    \caption{The electronics and data streaming setup is illustrated in a schematic diagram that includes the PYNQ board. The diagram shows how data is streamed through the board. The readout systems are managed by the Programmable Logic (PL) section and controlled via the \update{Processing System} (PS) section of the PYNQ board \citep[for more details, see][]{ele8358511}. The board's \update{Internet Protocol} (IP) address is used to operate the data acquisition process.}
    \label{fig:elec_schematics}
\end{figure*}

The electronics have three parts: a \stcb, the PYNQ\footnote{\updatethird{PYNQ\texttrademark\ is an open-source project from Advanced Micro Devices (AMD\textsuperscript{\textregistered}) that makes it easier to use Adaptive Computing platforms More information can be found at \url{https://www.pynq.io/}.}} board, and standard laboratory equipment like power supplies and a Personal Computer (PC) (Figure~\ref{fig:elec_schematics}).

\begin{figure*}[htb]
    \centering
    \includegraphics[width=0.85\textwidth]{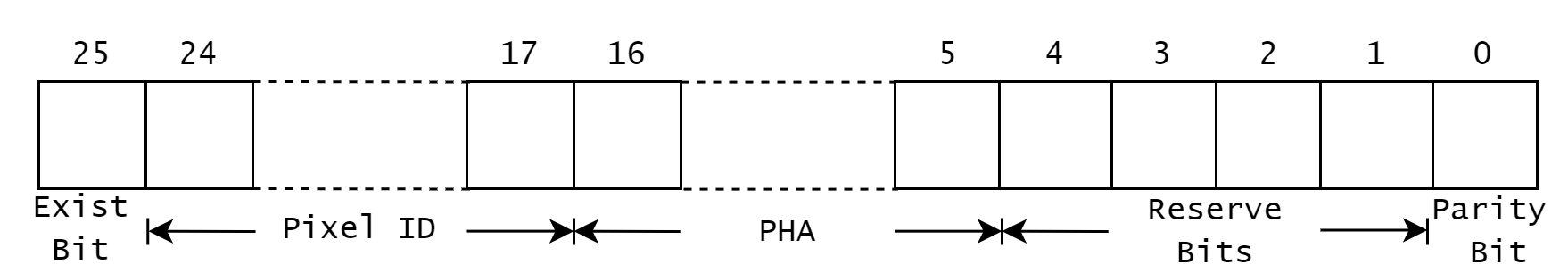}\\
    \includegraphics[width=1\textwidth]{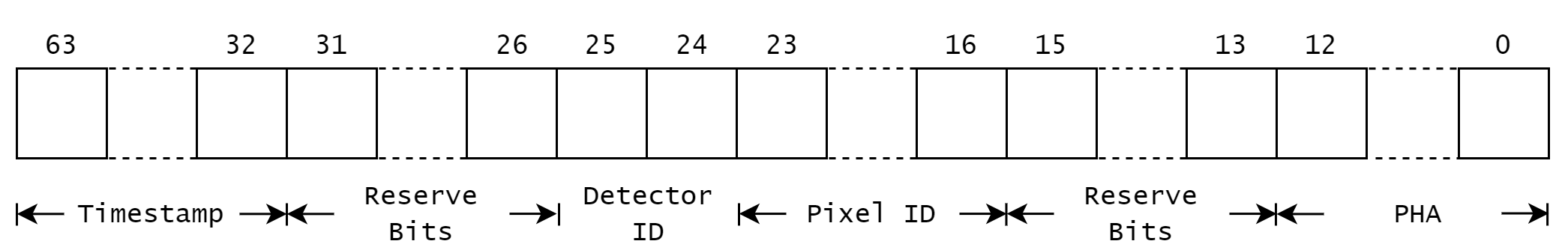}
    \caption{\textit{Top:} The native event data from the detector is 26-bits in length. The first 8 bits (Most Significant Bit, MSB) give the pixel number (\update{also known as the} ``channel address'') while the next 12 bits give the event energy. This is followed by 5 reserved bits, and the Least Significant Bit (LSB) is a parity bit. 
    \textit{Bottom:} A 32-bit timestamp at system clock resolution, a 2-bit detector ID and few reserve bits are added to the native data frame to make the 64-bit event data frame.}
    \label{fig:event_data_format}
\end{figure*}
The detectors are mounted on the \stcb, \updatethird{which has} the appropriate mounting pins for the high voltage and low voltage contacts of the detector, \updatethird{as well as SPI lines for data.} The board accepts 5~V input power, and converts it to the required 3.3~V and 1.8~V levels for the detector module. 

Data acquisition is handled by a Xilinx \update{Field Programmable Gate Array (FPGA)} PYNQ-Z2 board\footnote{\url{https://www.amd.com/en/corporate/university-program/aup-boards/pynq-z2.html}.}. It is a Zynq 7000 development board with ZYNQ XC7Z020-1CLG400C SoC (System on Chip). It has dual core ARM Cortex-A9 processors clocked at 650~MHz connected to various peripherals \update{including Ethernet}, as well as an Artix-7 equivalent FPGA. \update{A Peripheral Mode (Pmod) \update{Input-Output} (IO) Processor allows the FPGA to offload SPI communication to a dedicated microcontroller (typically an ARM Cortex-A9 inside the Zynq 7000 Processing System), improving efficiency and reducing FPGA fabric usage.} The SoC thus has two distinct hardware parts: the Processing System (PS) consisting of the ARM processors and its peripherals, and the Programmable Logic (PL). All the codes (C/C++ in standalone mode, or the Embedded Linux) run on the PS, while all the Custom IPs (software Intellectual Property cores) are created on the PL. The board also has 512~MB DDR3 memory with a 16-bit bus operating at 1050~Megabits per second. This is the main lower-speed buffer for all the data. Embedded Linux is installed on an SD card which accesses the RAM via the AXI--DMA (Advanced Extensible Interface, Direct Memory Access) Python code. 

We use small add-on circuits that convert the LVDS signal from the \stcb\ into the Low Voltage Transistor-Transistor Logic (LVTTL) levels used by the PYNQ board. A 10~MHz clock signal generated on the PYNQ board is supplied to the \stcb\ for clocking the detector. Detector data are read in the native 26-bit format as shown in Figure~\ref{fig:event_data_format} (top), which requires 7.5~$\mu$s time for each event, \update{thereby setting the native time resolution of our setup}. We add a 32-bit integer timestamp that counts the number of clock ticks that have elapsed since power-on. Our experiment does not require accurate absolute time, hence this suffices our requirements. This clock loops over in about 7 minutes, but this is fixed in post-processing of data by detecting the wrap-around and adding an appropriate offset. 

We split and pad the detector data and add the timestamp to save it in a 64-bit data format (Figure~\ref{fig:event_data_format}, bottom) for storing relevant photon information: the detector PHA\footnote{ PHA or Pulse Height Amplitude is the digital output corresponding to the signal measured by the circuit within the detector. PHA is measured in ``channels'', and is linearly related to the energy deposited in the pixel (\S\ref{subsec:gainoffset}).}, pixel number (PixID), detector number (DetID), and timestamp (number of clock cycles). 
 
We run Python Jupyter notebooks in the Embedded Linux on the SoC PS, and process the data with our custom codes (\S\ref{subsec:detinit}). For archival and further processing, data are read out from the PYNQ board over ethernet by a laboratory computer. The PYNQ board \update{requires a 7 to 12~volt stable supply} and can be powered via a USB cable from the computer, or from \update{external power supply provided by the vendor.} 
Lastly, the high voltage required by the detectors was supplied from a 
\update{CEAN Nuclear Instrumentation Module (NIM) Model N126 supply, with the total noise (Periodic and Random Deviation, PARD) $< 500~$mV, peak-to-peak.}

\subsection{Enclosure}\label{subsec:mech}

\begin{figure}[htb]
    \centering
    \includegraphics[width=1\columnwidth]{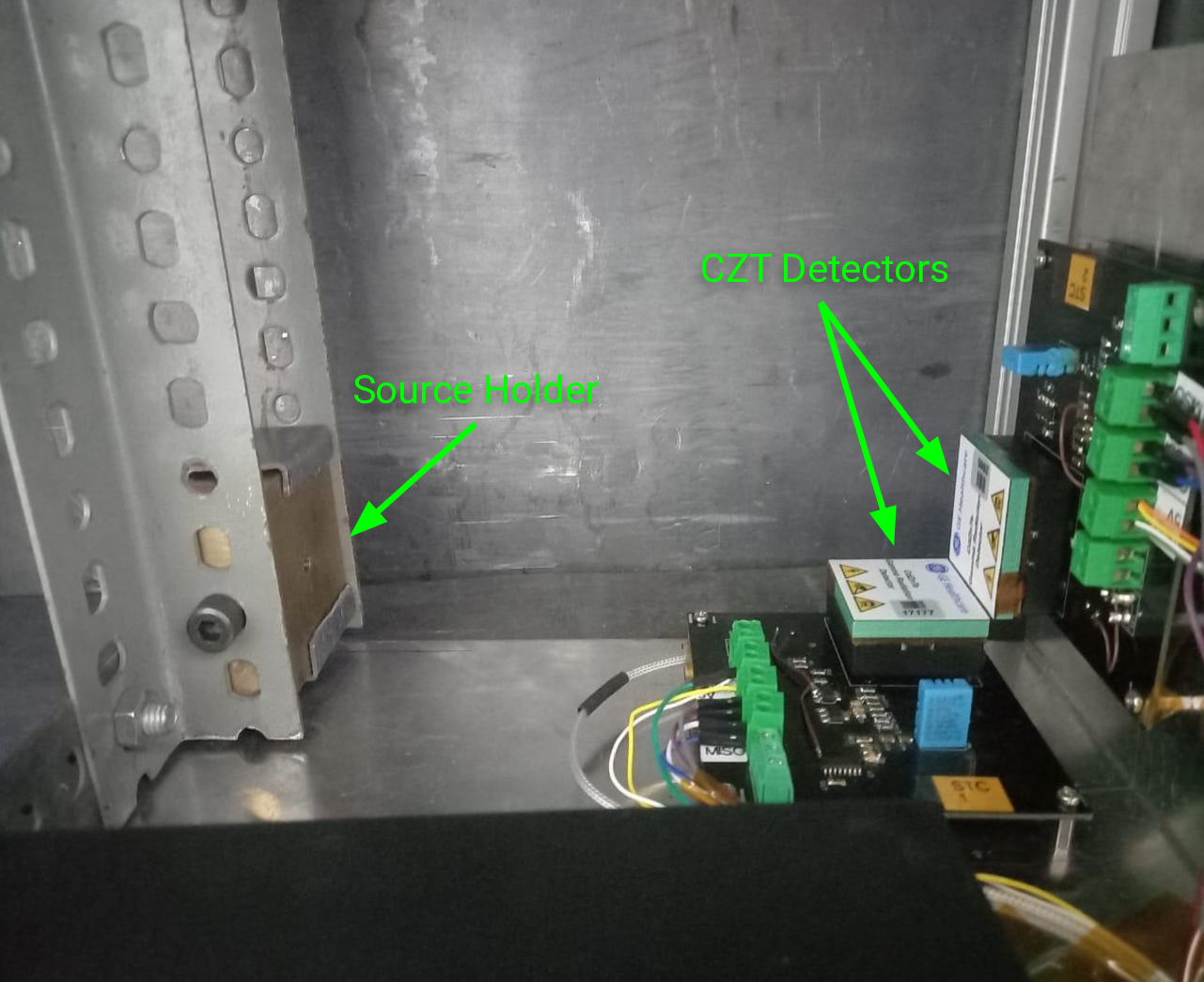}
    \caption{The experimental setup consisted of a holder containing a radioactive source to test the localization of hard X-rays at different positions and two perpendicular CZT detectors assembled on detector boards to stream the onboard data from the detector to the Xilinx FPGA controller for further data analysis.}
    \label{fig:setup_image}
\end{figure}

The detectors are sensitive to optical photons which can create noise events in edge pixels of the detector by creating electron-hole pairs in CZT, hence the entire setup has to be light-tight. As a first step, we do not put any indicator LEDs on any of the circuit boards. Next, we mount the detectors and the radioactive source in a 30~cm $\times$ 30~cm $\times$ 60~cm aluminium box, which is further covered with a black cloth. The box is lined with 5~mm lead sheets for safety of the operator. The detectors are mounted at right angles to each other, with a vertical detector (``Det~V'', detector ID~21065) serving as the scatterer and a horizontal detector (``Det~H'', detector ID~17177) serving as the absorber of scattered radiation (Figure~\ref{fig:setup_image}). 

Radioactive sources are enclosed in a thick brass shield with a cylindrical hole of 5~mm diameter and $24~$mm depth for collimating the radiation \update{such that most of the photons are incident on just the detector being used as a scatterer.} \updatetwo{We verify this by measuring the number of 81~keV photons detected in each detector, and find that Det~V receives 10 times more photons as compare to Det~H.} The source holder can be mounted at different heights so that photons are incident on Det~V at varied angles. 
Exact source location is measured mechanically to validate the source positions reconstructed from data.
\section{Data Acquisition and Calibration}\label{sec:data}

\subsection{Detector Initialisation} \label{subsec:detinit}
Like most \update{solid state detectors}, CZT detectors have lower noise when cooled. However, since it was not practical for us to cool the entire setup including the enclosure, we conducted our experiments at room temperature. This meant that we had more noise at low energies, and a higher number of noisy pixels in the detectors. As will be discussed in \S\ref{subsec:gainoffset}, we needed the ability to detect 31~keV photons from \element{133}{Ba}. Hence, we set the Lower Level Discriminator (LLD, also known as threshold) for both detectors to 20~keV.

Next, we disabled the noisy pixels in each detector. This was done iteratively by taking short acquisitions of background data and disabling noisiest pixels till a small subset of pixels no longer dominated the count rate. \update{We note that the set of noisy pixels can change each time we cycle power to the detector.} In most data acquisition ``runs'', typically less than ten pixels were disabled. The typical background count rate in each detector was finally around 8--10~\cps. If the detectors are cooled, this rate can drop down further.

\update{We found that under these conditions, the 31~keV line could be clearly detected in the spectra, as required.}

The highest PHA channel \update{(4095)} for the detector is used to denote any events with energy greater than or equal to the upper energy scale limit for the detector. Since exact energy information is not available for such events, we ignore them from our analysis.

\subsection{Gain-offset calibration}\label{subsec:gainoffset}
\begin{figure}[!htp]
    \centering
    \includegraphics[width=1\columnwidth]{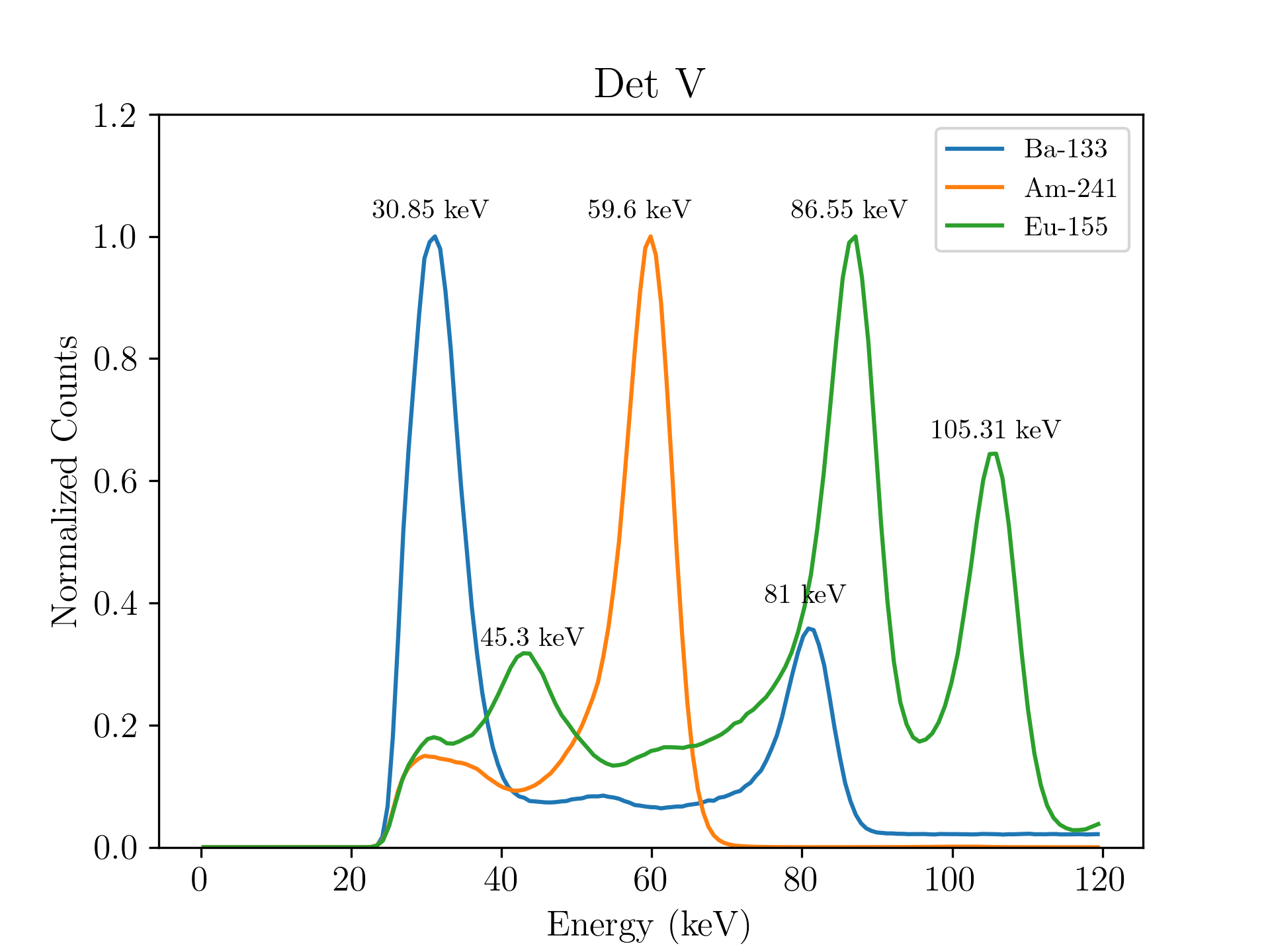}
    \includegraphics[width=1\columnwidth]{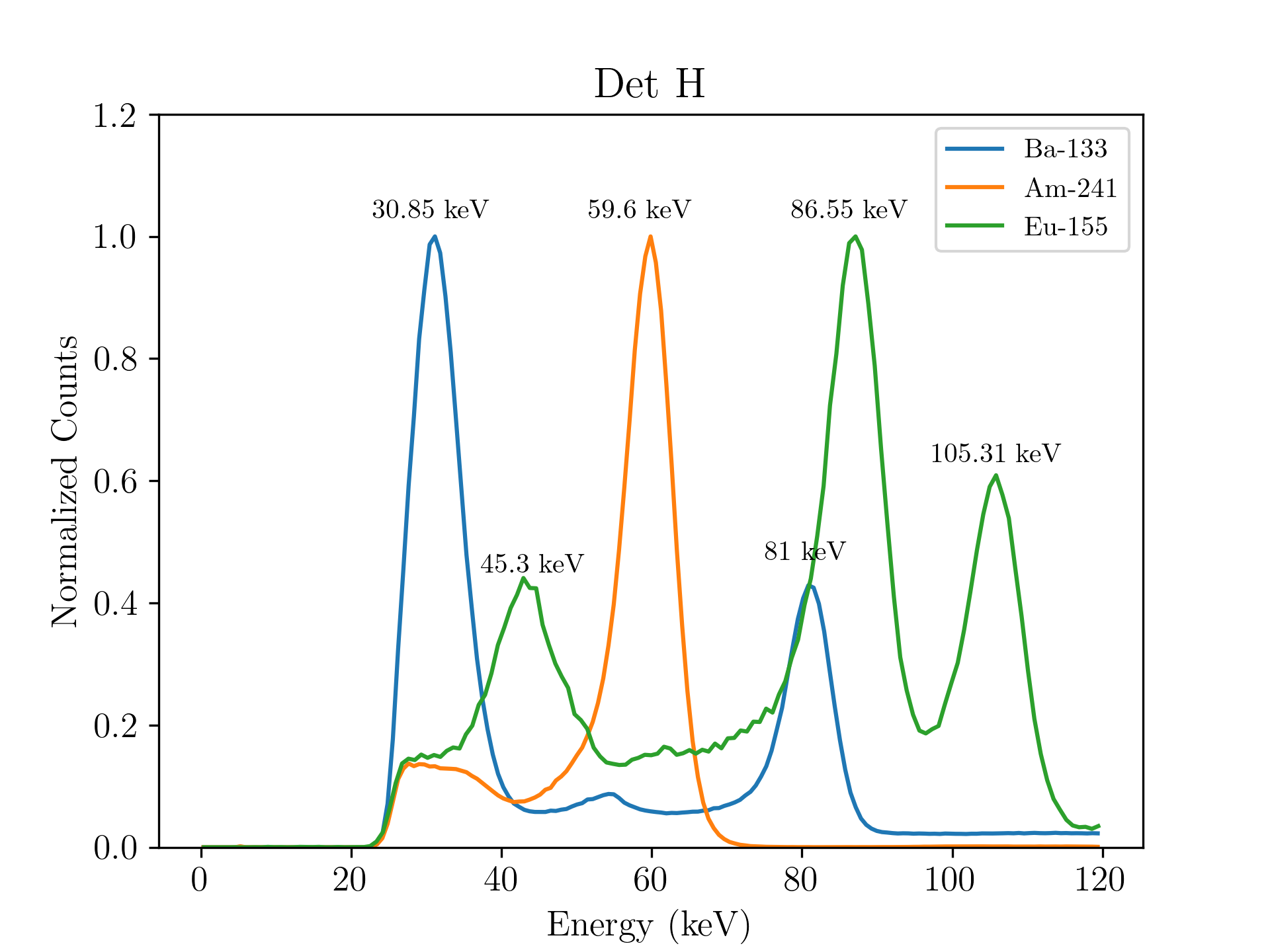}
     \caption{Spectra for various radioactive sources for both detectors. The spectra were separately acquired for each detector for each radioactive element, but have been plotted together for clarity. They are normalized such that the peak is at 1. The top and bottom figures are for \update{Det}~V and \update{Det}~H respectively. The green curves show  \element{155}{Eu} which produces three lines at 45.3~keV, 86.6~keV and 105.3~keV. The orange curve is for \element{241}{Am}, showing a line at 59.6~keV. Lastly, blue curves show \element{133}{Ba} data with lines at 30.9~keV and 81~keV. Note that the 31~keV line is close to the detector LLD, and is likely impacted by the cutoff as well as electronic noise.}
\label{fig:ba_peaks_on_dets}
\end{figure}

Lab measurements show that the PHA measured by the detector is linearly related to the energy ($E$) of the incident photon for these detectors. We model this as $E = G\times \mathrm{PHA} + O$ where $G$ is the detector gain, and $O$ is the offset in units of energy. The values of $G$ and $O$ \updatetwo{vary from pixel to pixel, and need to be measured explicitly for the setup}.
We irradiate the individual detectors sequentially with \element{133}{Ba}, \element{155}{Eu}, \element{241}{Am}, which have multiple lines \citep{kinsey1996nudat} \footnote{Available isotope--wise at \url{https://www.nndc.bnl.gov/nudat3/decaysearchdirect.jsp?nuc=133Ba&unc=NDS}} in the 20--200~keV range of interest (Figure~\ref{fig:ba_peaks_on_dets}). \update{While the 31~keV line is detected in data as required, we find that the strongly non-linear shape of the background results in a shift in the measured gaussian peak. Since multiple other lines are cleanly detected (Figure~\ref{fig:ba_peaks_on_dets}), we do not use the 31~keV line for gain-offset measurements.}
We use pixel--wise spectra for each source, measure the PHA of the peak, and then calculate gain and offset values. These values are then tabulated and used in all subsequent analyses.
To calculate the energy resolution of the detectors, we irradiate the detectors with \element{241}{Am} source on them, and create an average detector spectrum (Figure~\ref{fig:ba_peaks_on_dets}) by \update{converting each photon's PHA into energy by using the gain and offset values of the corresponding pixel.} We measure an energy 
resolution of $\sim12~\%$ at 59.6~keV at room temperature for both detectors, on par with values obtained in \asat/CZTI \citep{Bhalerao2017}.
\subsection{Data collection at various source positions}
Final experimental data sets were obtained by placing the \element{133}{Ba} source at four different locations in the enclosure. \element{133}{Ba} has a prominent radioactive decay lines at 356~keV and 81~keV, with decay yields of 62\%\ and 34\%\ respectively. The source also has weaker lines at \update{31~keV,} 276~keV, 303~keV, 384~keV, etc. Of all these, the 356~keV is most suitable for our analysis due its high yield, relatively high probability of Compton scattering \update{as compared to photoelectric absorption}, and because the energy deposited in both the scatterer and \update{the} absorber is likely to be within our detector energy range. The 2.5~$\mu$Ci source strength corresponds to about 9.25\ee{4} disintegrations per second, of which a much smaller fraction exits through the collimator. We typically detect 35--65~\cps\ on our detector. \update{This number} includes the 31 and 81~keV lines from \element{133}{Ba}. Photons of 356~keV which are directly absorbed in the detector \update{get a PHA of 4095, and} get discarded as discussed in \S\ref{subsec:detinit}. However, if these photons are Compton scattered by the detector and the energy deposit is in the 20--200~keV range, such events will be registered in the scatterer (Det~V). The \update{photons are scattered in arbitrary directions}, and a fraction of them are incident on the absorber (Det~H). If the energy of the scattered photon is also in the 20--200~keV range, then these register as events in Det~H.
 
The final count rates were typically around 35--65~\cps\ in Det~V depending on the source position, and around 15~\cps\ in Det~H respectively. This includes background rates of 8--10~\cps\ in each detector. We find that a few detector pixels occasionally become noisy during the observations, increasing the count rates to as much as few hundred counts per second. These events are stored along with the data, but we discard all data from such pixels in post-processing. In our test runs, we typically acquired data for 4--5 hours to get sufficient statistics for analysis.

\section{Compton Event Reconstruction}\label{sec:reconstruct}
There are various methods to infer the source position and other characteristics in a Compton camera starting from simple Back--Projection of event circles onto the sky to iterative imaging reconstruction approaches like Maximum-Likelihood Expectation-Maximization (MLEM) algorithm and the Maximum Entropy Method (MEM) \citep{zoglauerthesis,Frandes2016}. Our focus in this work is the demonstration of a Compton camera in sub--MeV range, hence in our analysis \update{we start with the} straight forward back--projection method, which we discuss in this section. \update{Then, we reconstruct source positions with the MLEM method (\S\ref{ref:mlem}).} We discuss some of the limitations for this in \S\ref{subsec:limitations}.

\subsection{Event circles and source position}\label{sec:evt_circles}
\begin{figure*}[!htp]
    \centering
    \includegraphics[height=3in]{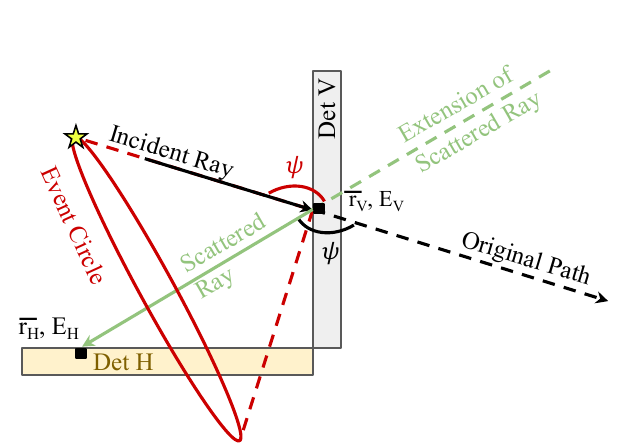}\\
    \includegraphics[height=3in]{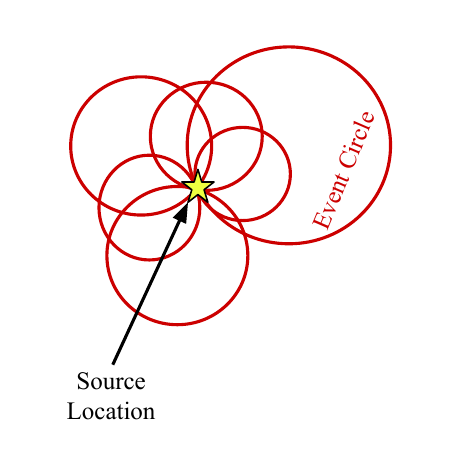}    
    \includegraphics[height=3in]{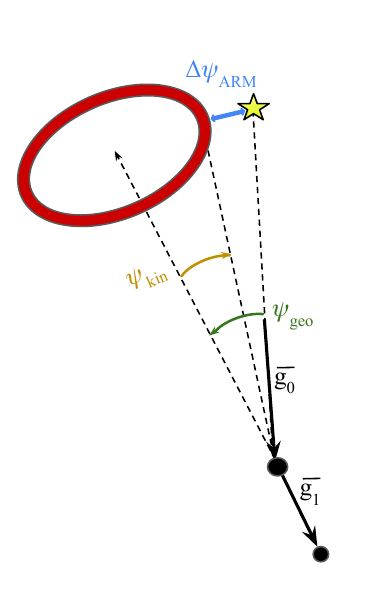}
      \caption{\textit{Top:} A incident photon hits Det~V at position $\overline{r_V}$, deposits part of its energy as $E_V$, scatters off and then hits Det~H at position $\overline{r_H}$, depositing its remaining energy as $E_H$ via photoelectric absorption. The scattering angle $\psi$ determines the half angle of the event circle cone (obtuse scattering angle reverses the cone's direction). \textit{Bottom left:} When multiple event circles are back-projected on the sky, the common overlap position determines the location of the source, as is visualized for six event circles. \textit{Bottom right:} The geometrically measured Compton scattering angle is given by the angle between the initial photon direction $\overline{g_0}$ and the scattered gamma-ray direction $\overline{g_1}$. The difference between the kinematic and geometric Compton scatter angle gives the Angular Resolution Measure (ARM), $\Delta \varphi_{ARM}$ for each event. Based on a figure in \citet{kierans2022}.}
    \label{fig:compton_imaging_working}
\end{figure*}

Consider a photon incident on Det~V at a position $\overline{r_V}$, where it undergoes Compton scattering in Det~V, depositing energy $E_V$ at the point of incidence (Figure~\ref{fig:compton_imaging_working}, top). The scattered photon lands on Det~H at position $\overline{r_H}$ and is fully absorbed, depositing its entire energy $E_H$ at that position. One can calculate the scattering angle $\psi$ by applying energy and momentum conservation, leading to the well-known Compton equation:
\begin{equation}
    \cos \psi = 1 - \frac{m_ec^2}{E_H} + \frac{m_ec^2}{E_V+E_H} \quad .\label{eq:scat_angle}
\end{equation}
Combining this with the scattering positions $\overline{r_V}$ and $\overline{r_H}$, we can constrain the source location to a cone (Figure~\ref{fig:compton_imaging_working}, top). The intersection of this cone with a large spherical surface (the ``sky'') centred on the detector gives a circle known as the Event Circle. The intersection of circles created by multiple events gives us the location of the source in the sky (Figure~\ref{fig:compton_imaging_working}, lower left).

In practice, the event circles have finite widths, due to uncertainties in measurements. We account for these by modeling the event circle as a Gaussian band on the sky.

The first key factor is the ``energy uncertainty'' ($\Delta\psi_E$) stemming from the finite energy resolution of the detector, given by the following equation \citep{zoglauerthesis}: 
\begin{align}
\Delta\psi_{E}  = \frac{m_ec^2}{\sin\psi}
     & \Bigg[ \left(\frac{1}{E_H^2}-\frac{1}{(E_H+E_V)^2}\right)^2 \Delta E_H^2 \nonumber \\
     & + \left(\frac{1}{E_H+E_V }\right)^4 \Delta E_V^2 \Bigg]^{1/2}. 
     \label{eq:delta_psi_E}   
\end{align}

The second key factor is the positional uncertainty ($\Delta \psi_P$) which arises from the 2.46~mm pixel size, which is not negligible as compared to the few cm length that the scattered photon travels: resulting in an uncertainty in the cone axis \update{$\overline{r_V} - \overline{r_H}$}. For this calculation, we use the average uncertainty corresponding to a pair of pixels near the center of the horizontal and vertical detectors, $\Delta \psi_P = 2.53\degr$ (one-sigma). For each photon, we calculate the net uncertainty ($\Delta \psi$) by adding these terms in quadrature. The event circle is then constructed as a Gaussian with standard deviation equal to $\Delta \psi$ and truncated at $\pm 3\sigma$.
\subsection{Compton Event Selection}\label{sec:compsel}
Our \update{clock speed} is 0.1~$\mu$s, while the light travel time for a scattered photon between the two detectors is sub-nanosecond. Hence the simple criterion for selecting Compton events is that photons should have the same timestamp in both detectors. We note that our codes on the PYNQ board PL run in parallel acquiring data from both detectors simultaneously, thereby ensuring that such photon pairs would indeed get the same timestamp. 

\begin{figure*}[thbp]
    \centering
    \includegraphics[width=0.8\textwidth]{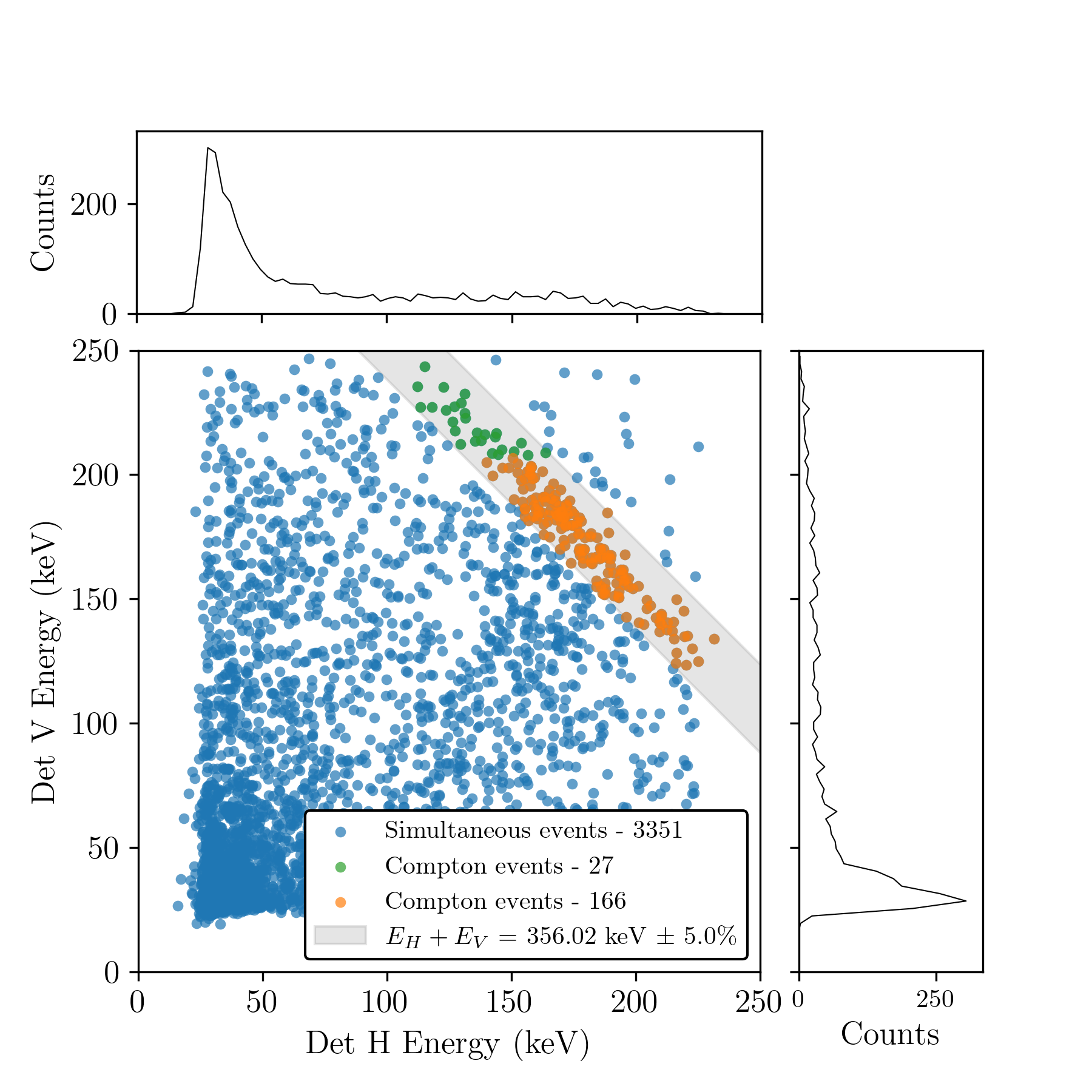}
    \caption{3351 Compton event pairs detected in a test run. The scatter plot shows the energy deposited in Det~V and Det~H respectively. The gray band denotes the region of the plot where we expected to get 356~keV \element{133}{Ba} photons, with a width set to the $\pm 5\%$ energy resolution of that line. All blue dots therefore denote chance coincidence events. The orange and green dots denote events where the scatterer was considered to Det~V and Det~H respectively. 
    The plots at the top and right show spectra of Compton events in Det~H and Det~V respectively. We can see the low-energy peaks in both, which correspond to electronic noise. Note that the 81~keV line is not clearly seen in these spectra.}
    \label{fig:e0_vs_e1}
\end{figure*}

All such event pairs from one of our data sets are shown in Figure~\ref{fig:e0_vs_e1}. The histograms on the sides show the spectra for such events in the individual detectors. We see that the individual detector spectra show a peak at low energies, consistent with being dominated by electronic noise.

\begin{figure}[thbp]
    \centering
    \includegraphics[width=1\columnwidth]{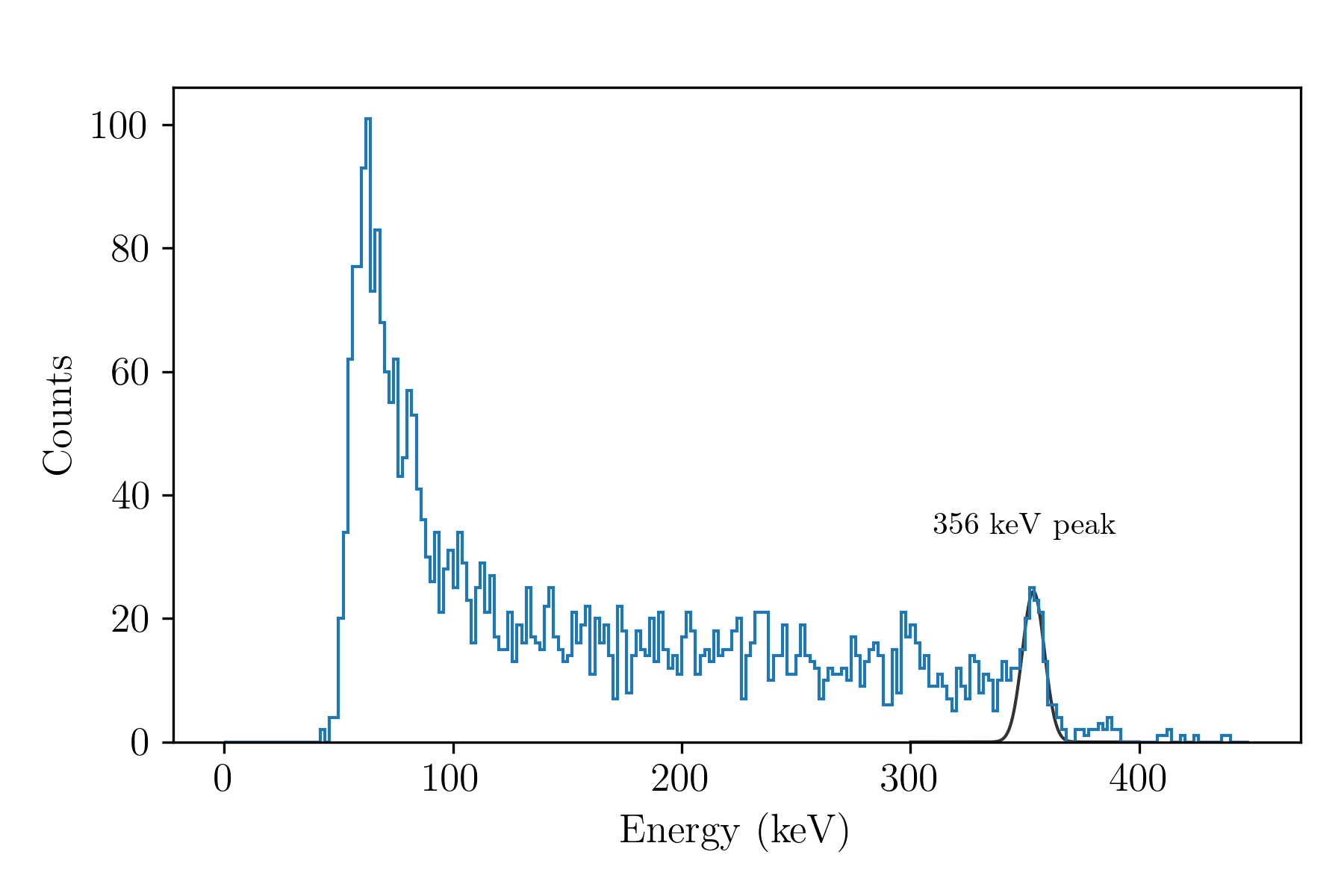}
    \caption{Compton event spectrum from the test run shown in Figure~\ref{fig:e0_vs_e1}. This spectrum is created by using the total energy of photons in each Compton pair, without applying any selection for the 356~keV line. We see a small peak at the expected line position, and a prominent drop on its high energy side. A small number of photons seen at higher energies may be from the 384~keV \element{133}{Ba} line, and there are also hints of the 303~keV line. Fitting the 356~keV line profile with a Gaussian gives a resolution of 3.2\%\ (FWHM).}
    \label{fig:ev_plus_eh_hist}
\end{figure}

Since our background noise is relatively higher, we need another filter to limit our data to true Compton event pairs. We expect that the \element{133}{Ba} source photons undergoing Compton scattering must all be from the 356~keV decay line. By constructing a spectrum of the total deposited energy ($E_V + E_H$), we can detect the 356~keV line (Figure~\ref{fig:ev_plus_eh_hist}). We see a prominent drop on the high energy side of the line, indicating an absence of higher energy photons from the source. Fitting the line profile with a Gaussian, we measure an energy resolution of 3.2\% (FWHM). As a cross-check, we divided the event data into two halves, and measured the line energy from both. We find that the measurements agree to better than 0.1~keV.
To select photons from this line for our analysis, we apply a conservative filter, selecting all photon pairs where the total energy is within $\pm 5\%$ of the line energy: $ |E_V + E_H - 356~\mathrm{keV}| < 17.8$~keV.
 
The source is collimated such that photons should be incident on Det~V. However the collimation is not perfect and some photons may be incident on Det~H. There is no fool--proof way to determine the scatterer and absorber for a given Compton photon pair at our time resolution. Instead, we use the concept of the Compton edge: the maximum energy that can be deposited in a scatterer for a source photon of a given energy \citep[Equation~20]{kierans2022}. For 356~keV photons from \element{133}{Ba}, this corresponds to 206~keV. Hence, if we find a Compton photon pair where $E_V > 206$~keV, we conclude that the photon must have been scattered from Det~H and absorbed in Det~V. Such events are shown as green dots in Figure~\ref{fig:e0_vs_e1}, while orange dots show events where we have considered Det~V to be the scatterer.

\subsection{Source position by back-projection}
\begin{table}[htp]
    \centering
    \begin{tabular}{ccc}
    \hline
      Serial no. & Polar Angle  & Azimuthal Angle \\
                 & ($\theta$, deg) & ($\phi$, deg) \\   
      \hline
        1 &  $-3$\degr & 173\degr \\
        2 & 20\degr & 184\degr \\
        3 & 38\degr & 184\degr \\
        4 & 24\degr & 190\degr \\
      \hline
    \end{tabular}
    \caption{\element{133}{Ba} sources were placed at predefined positions in four different cases. For these measurements we used a coordinate system origin at the centre of the surface of Det~V. In Figure~\ref{fig:skymaps}, these source positions are marked with a bold cross mark.}
    \label{observation table}
\end{table}

\begin{figure*}[htp]
    \centering
    \includegraphics[width=0.45\textwidth]{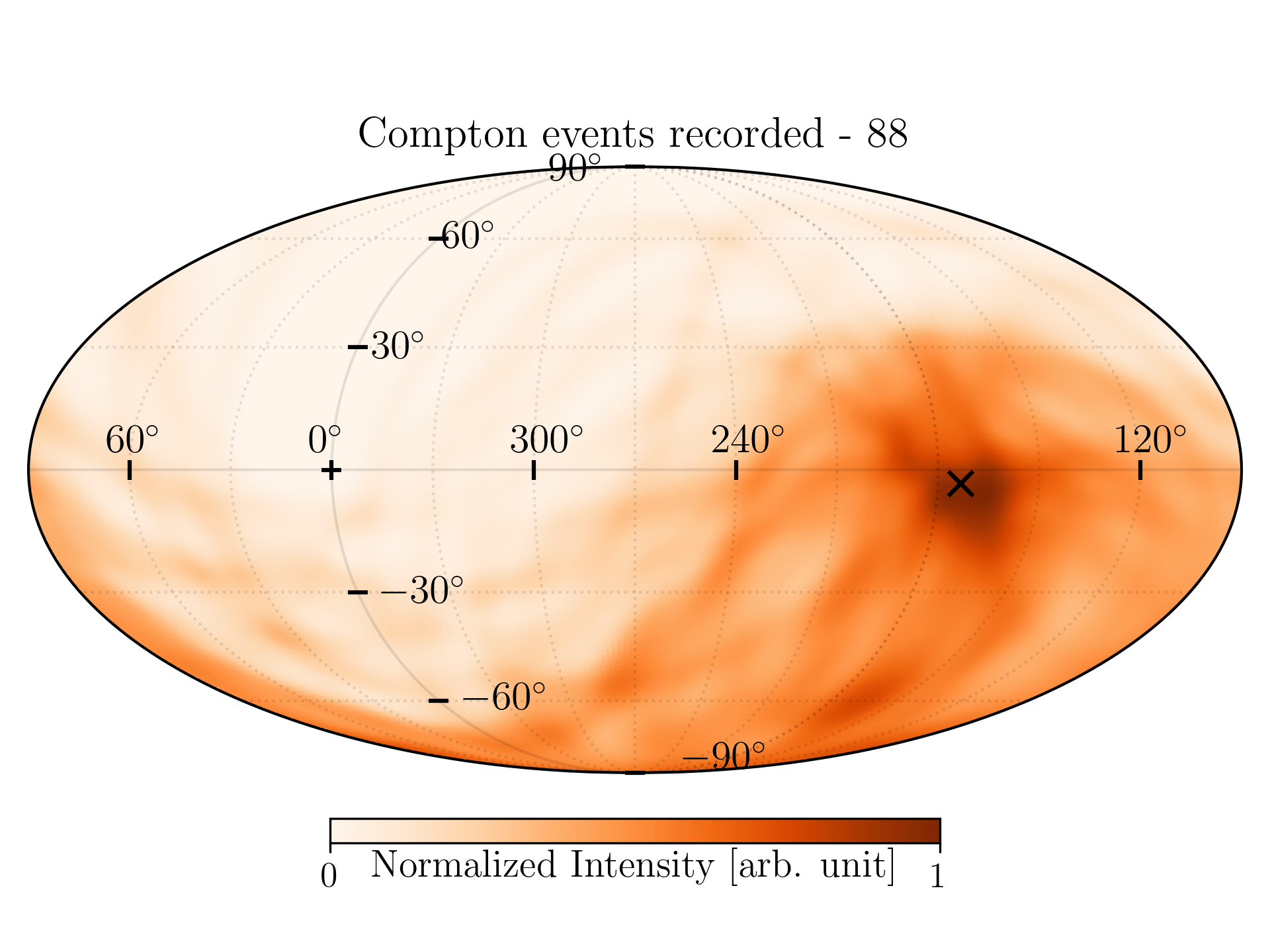} \hfill
    \includegraphics[width=0.45\textwidth]{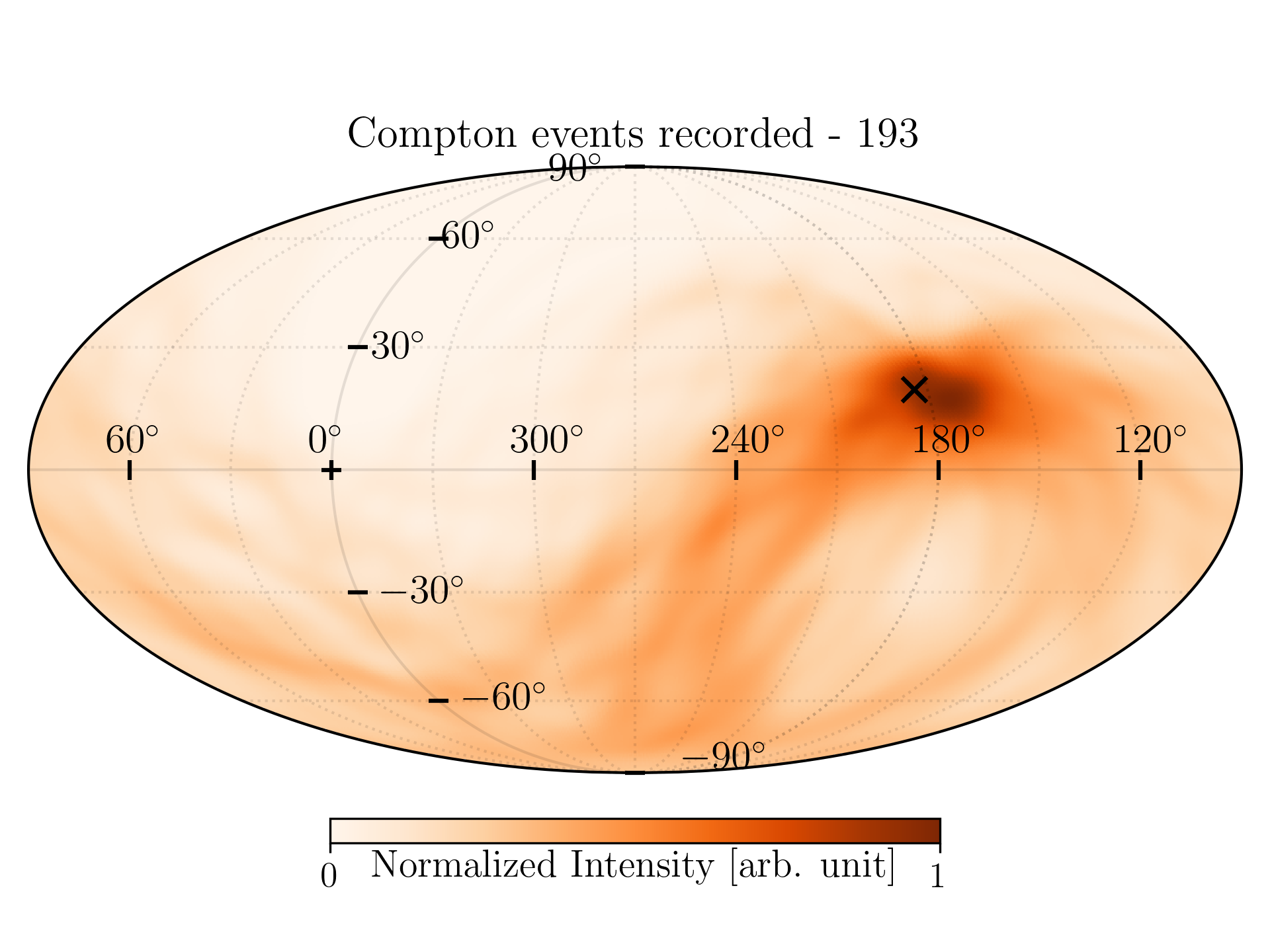}\\
    \includegraphics[width=0.45\textwidth]{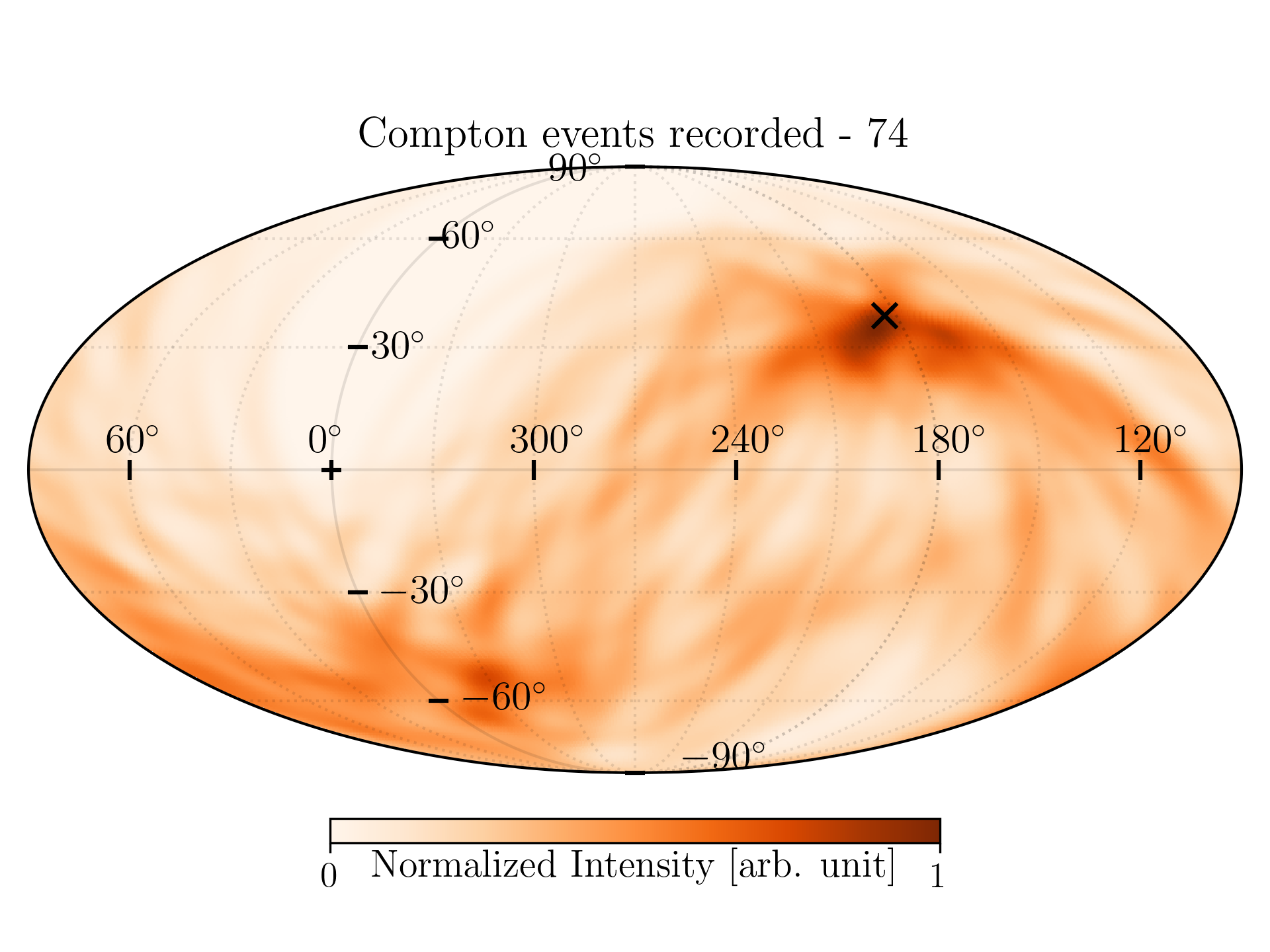} \hfill
    \includegraphics[width=0.45\textwidth]{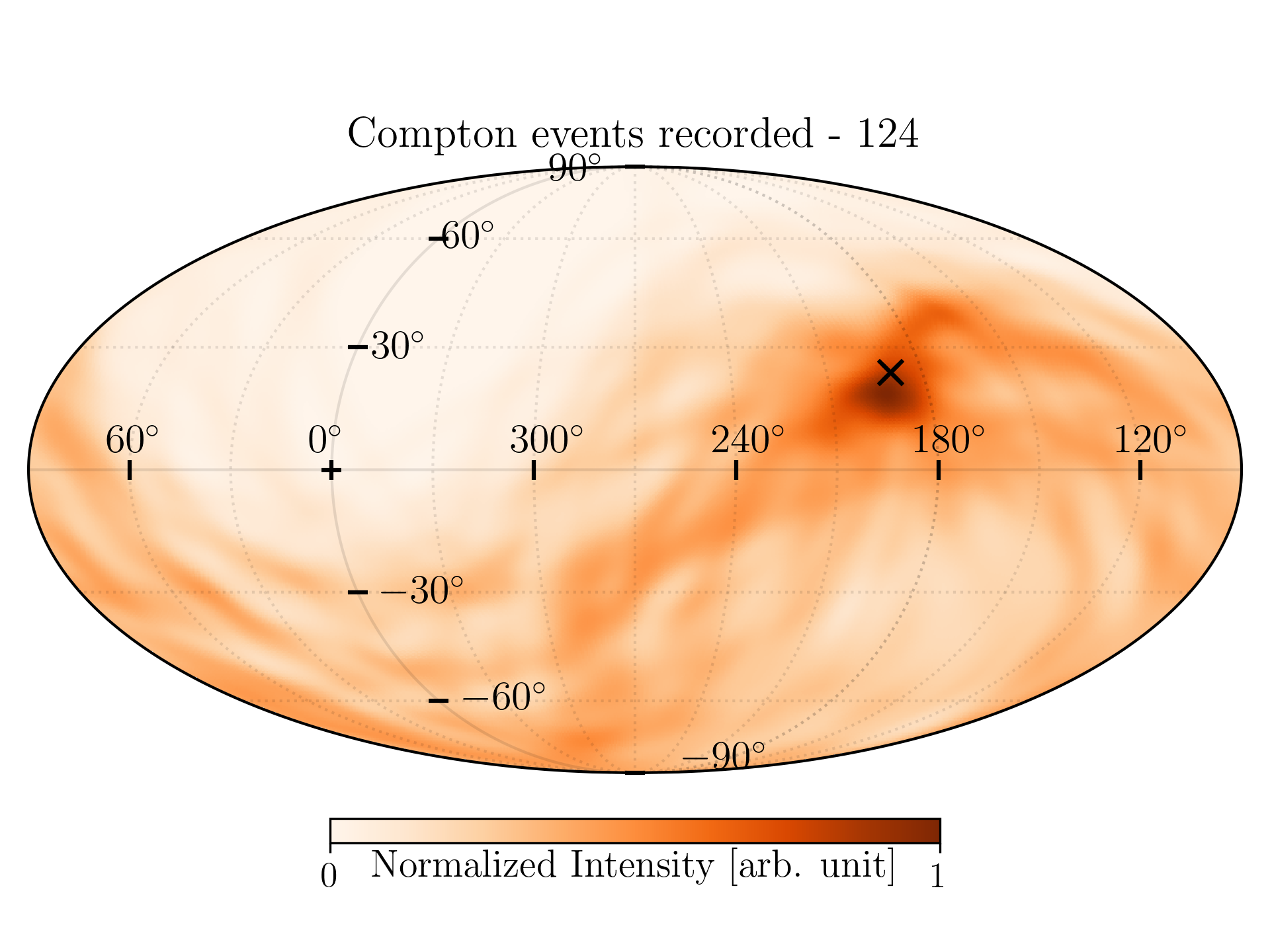}
     \caption{\update{Source position reconstruction using back-projection method.} The all-sky maps show the event circles with a finite width based on the detector's energy and position resolution uncertainties. These circles overlap, forming a high-probability region where the source is likely to exist. The actual source location (marked by a cross) was found to be within these regions.
    \textit{Upper left:} Case 1 ($\theta = -3\degr, \phi=173\degr$). \textit{Upper right:} Case 2 ($\theta = 20\degr, \phi=184\degr$). \textit{Lower Left:}  Case 3 ($\theta = 38\degr, \phi=184\degr$). \textit{Lower Right:} Case 4 ($\theta = 24\degr, \phi=190\degr$).}
    \label{fig:skymaps}
\end{figure*}

The results were obtained using four different coordinates of the \element{133}{Ba} source, as listed in observation Table~\ref{observation table}. We then use the back-projection algorithm and map each event to a ring on the sky as discussed in \S\ref{sec:evt_circles}. The final sky maps are created by adding all event rings from the selected 356~keV events. The source location corresponds to the high probability region resulting from this process (Figure~\ref{fig:skymaps}). We can see that in all four cases, our reconstructed position is consistent with the true source position (marked with a bold cross). \update{We note that the collimator is 5~mm wide, and placed at distances ranging from 22~cm to 28~cm from the scatterer. Thus the source appears about 1\degr -- 1.3\degr\ in angular size as seen from the scatterer, which is about the thickness of the lines of the cross.}

Given the knowledge of the true source position, we can now evaluate the angular resolution of our setup. The best-fit position deviates from the true position due to various factors like the native angular resolution of the setup, geometric measurement uncertainties, and simplifying assumptions made in our analysis. We now evaluate the Angular Resolution Measure (ARM) of our setup, and discuss some of the other limitations in \S\ref{subsec:limitations}.
 
The angular resolution of a Compton camera is defined as the full width at half maximum (FWHM) of the ARM distribution \citep{kierans2022}. For each Compton pair, we can calculate the Compton scatter angle $\psi$ in two ways. The geometric Compton scatter angle $\psi^{geo}$ is calculated from the actual positions of the source, scattering pixel, and absorbing pixel (Figure~\ref{fig:compton_imaging_working}) as a dot product of the initial photon direction $\overline{g_0}$ and the measured scattered gamma-ray direction {$\overline{g_1}$}. The kinematic Compton scatter angle ($\psi^{kin}$) is determined from the measured energy deposits using Equation~\ref{eq:scat_angle}, and gives us a ring on the sky. The ARM for each event is defined as the minimum distance between the ring and the geometrically calculated source direction (Figure~\ref{fig:compton_imaging_working}, lower right):
\begin{equation}
    \Delta\psi_{\text{ARM}} = \psi^{geo} - \psi^{kin} \quad .
\end{equation}

\begin{figure*}[htp]
    \centering
    \includegraphics[width=1\columnwidth]{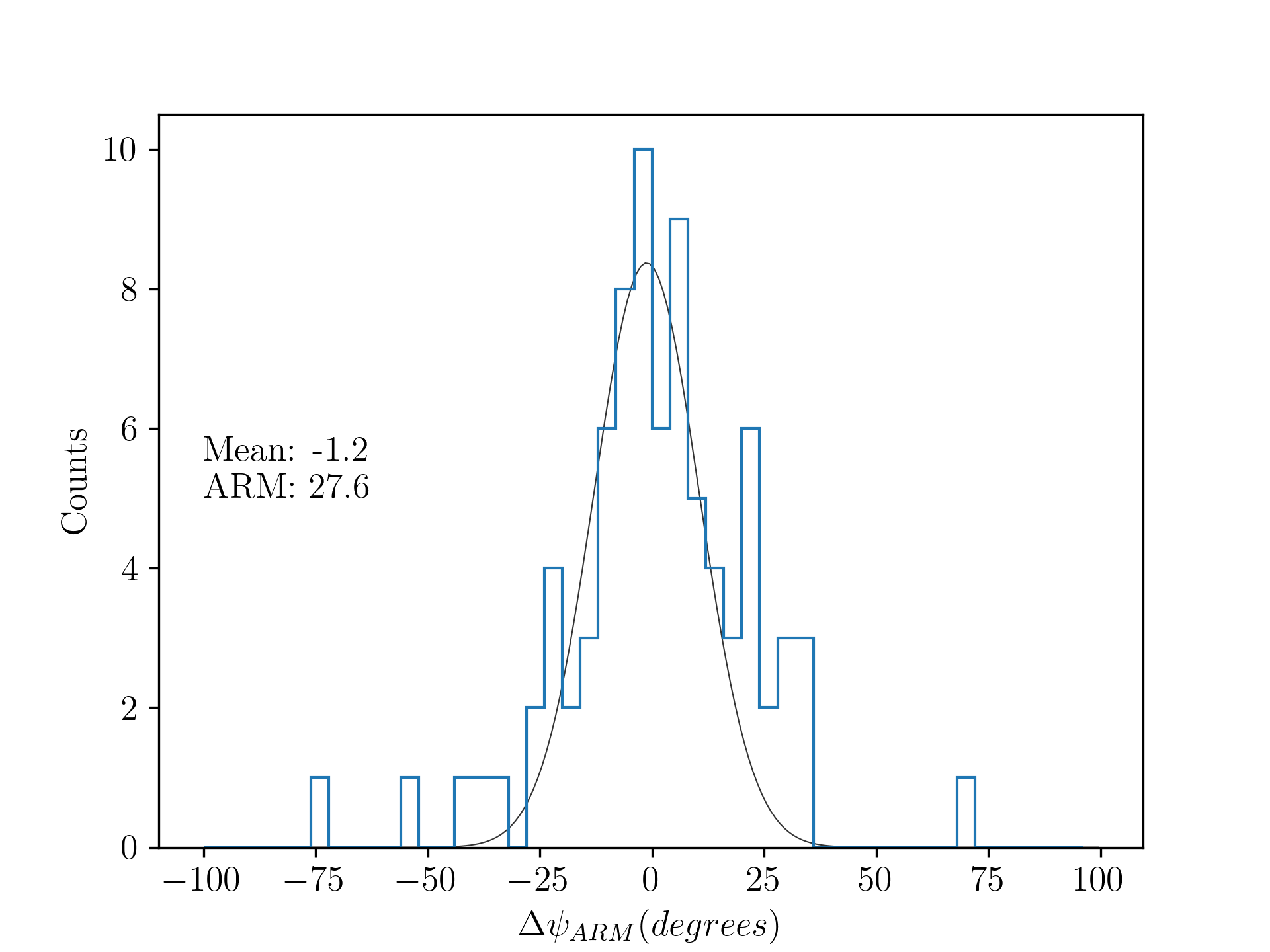} \hfill
    \includegraphics[width=1\columnwidth]{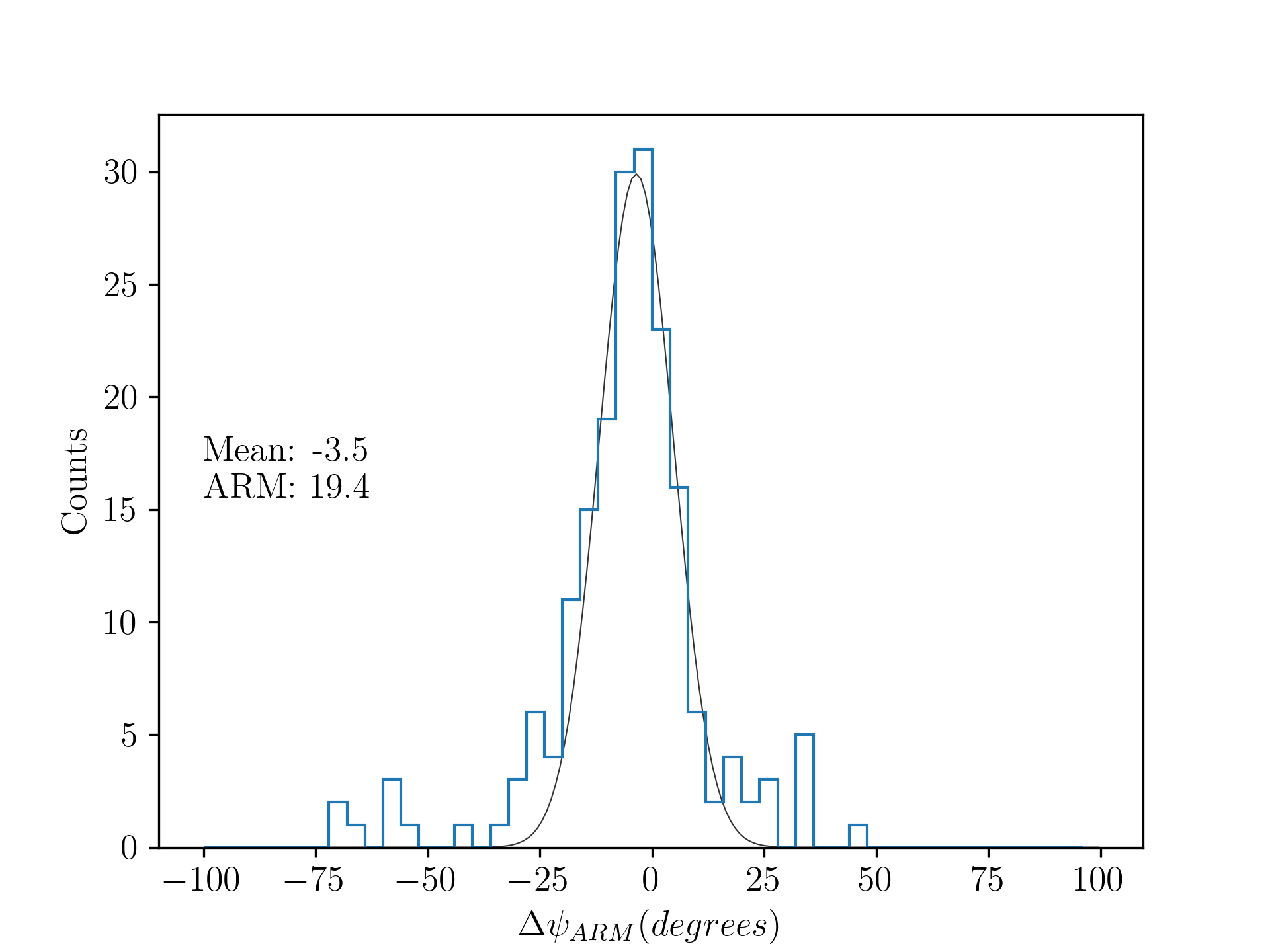}\\
    \includegraphics[width=1\columnwidth]{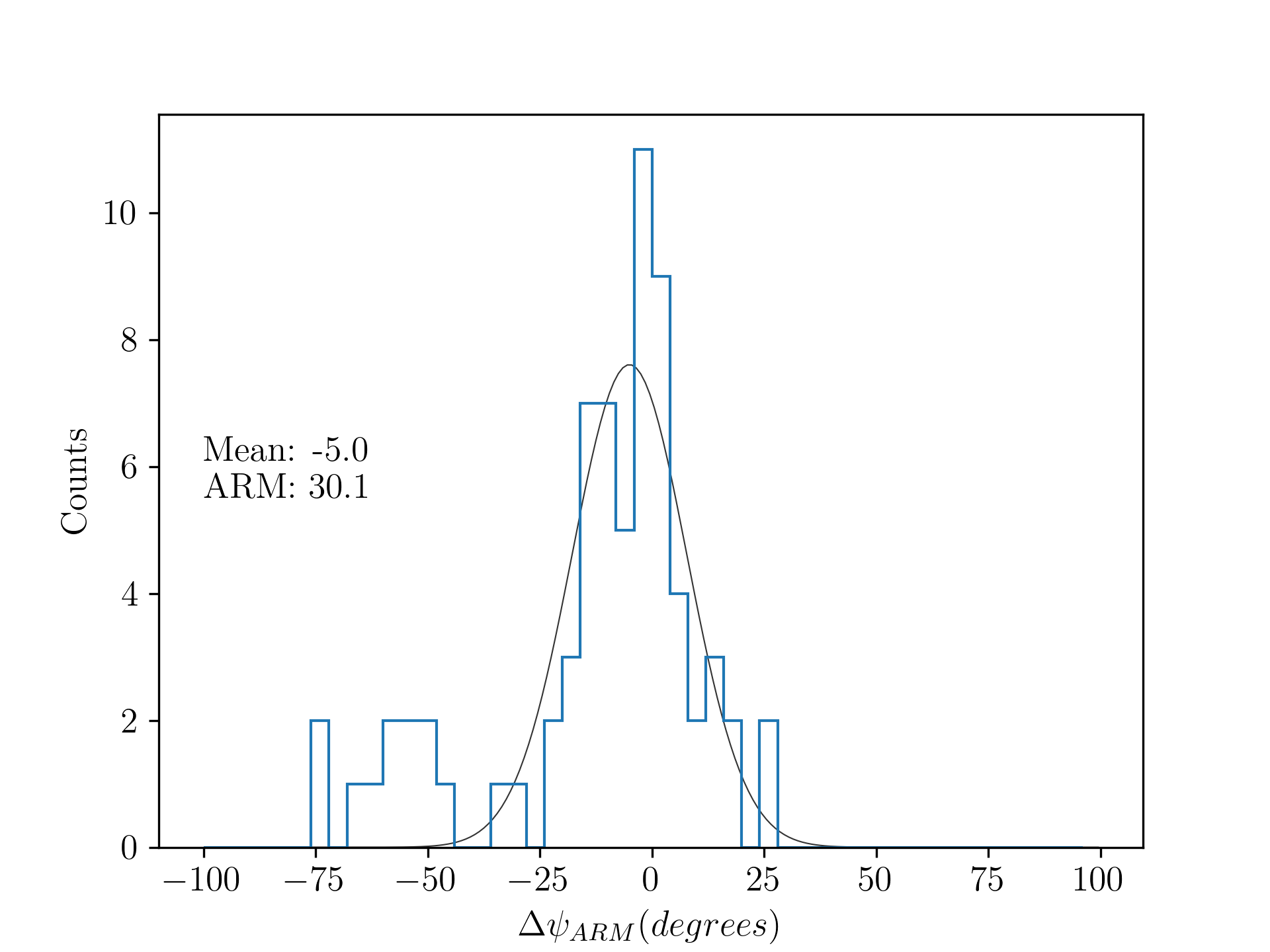} \hfill
    \includegraphics[width=1\columnwidth]{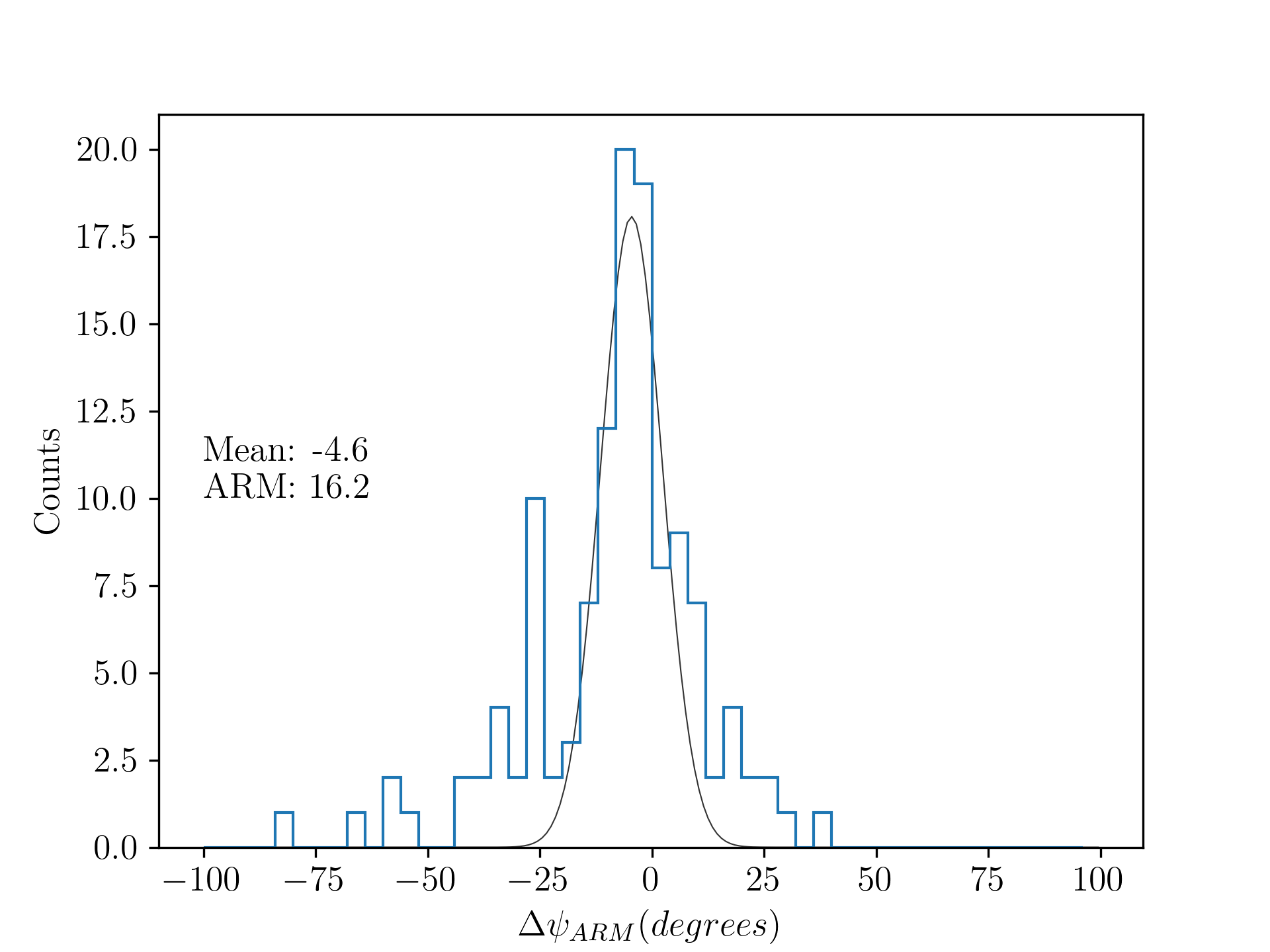}
    \caption{A histogram of the $\Delta\varphi_{\text{ARM}}$ values for the four tests from Table~\ref{observation table}. 
    \textit{Upper left:} Case 1 ($\theta = -3\degr, \phi=173\degr$). \textit{Upper right:} Case 2 ($\theta = 20\degr, \phi=184\degr$). \textit{Lower Left:}  Case 3 ($\theta = 38\degr, \phi=184\degr$). \textit{Lower Right:} Case 4 ($\theta = 24\degr, \phi=190\degr$).
    A Gaussian fit to the main peak in each histogram gives the FWHM of the ARM, corresponding to the resolution of the Compton imaging system. As expected for our setup, we find that the value varies with direction. We obtain ARM in the range from $\sim 16\degr$ to $\sim 30\degr$.}
        \label{fig:arm_plots}
\end{figure*}

The histogram of $\Delta\psi_{\text{ARM}}$ values is shown in Figure~\ref{fig:arm_plots} for our four experiments. For an ideal detector, we expect a sharp peak at $\Delta\psi_{\text{ARM}}=0$, with a very narrow width. In practice, we expect to see a distribution centered close to zero. The FWHM of this distribution is the angular resolution of the Compton Imager. We see that our setup has a direction--dependent energy resolution, ranging from $\sim 16\degr$ to $\sim 30\degr$.

\subsection{Maximum-Likelihood Expectation-Maximization}\label{ref:mlem}
\update{Now, we apply the MLEM algorithm \citep{wilderman1998list, yabu2021tomographic} to localize the reconstructed source with better noise suppression to obtain a sharper localization. This algorithm iteratively refines the source intensity distribution by maximizing the likelihood of observed Compton events given the system response. The sky is divided into pixels and the image is expressed as an intensity $\lambda_j$ for each pixel $j$. The image $\lambda_{j}^{k+1}$ after $k+1$-th iteration is calculated from the $k$-th image $\lambda_{j}^{k}$ as:}
\begin{equation}
    \lambda_{j}^{k+1} = \frac{\lambda_{j}^{k}}{s_{j}} \sum_{i}\frac{t_{ij}}{\sum_{l}t_{il}\lambda_{l}^{k}} \quad ,
\end{equation}
\update{where
\begin{itemize}
    \item $i$ represents the index of a Compton event, 
    \item $j$ represents the index of the sky pixel, 
    \item $s_j$ is the probability that a photon emitted from pixel $j$ would be detected as a Compton event,
    \item $t_{ij}$ is the probability that a photon emitted from pixel $j$ creates the Compton event $i$,
    \item \updatetwo{the variable $l$ denotes individual pixels and is used as a summation variable to sum over all pixels in the denominator for each $i$, $k$.}
\end{itemize}}

\update{The value of $s_j$ does not impact the inferred distribution of sources on the sky, but serves as a normalisation factor. Since our interest is only in the source position, we can ignore the overall normalisation and simply use $s_j =1$. $t_{ij}$ is calculated using the sky map generated for each event circle $i$ for a given Compton event as discussed in \S\ref{sec:evt_circles}.}

\begin{figure*}[htp]
    \centering
    \includegraphics[width=0.45\textwidth]{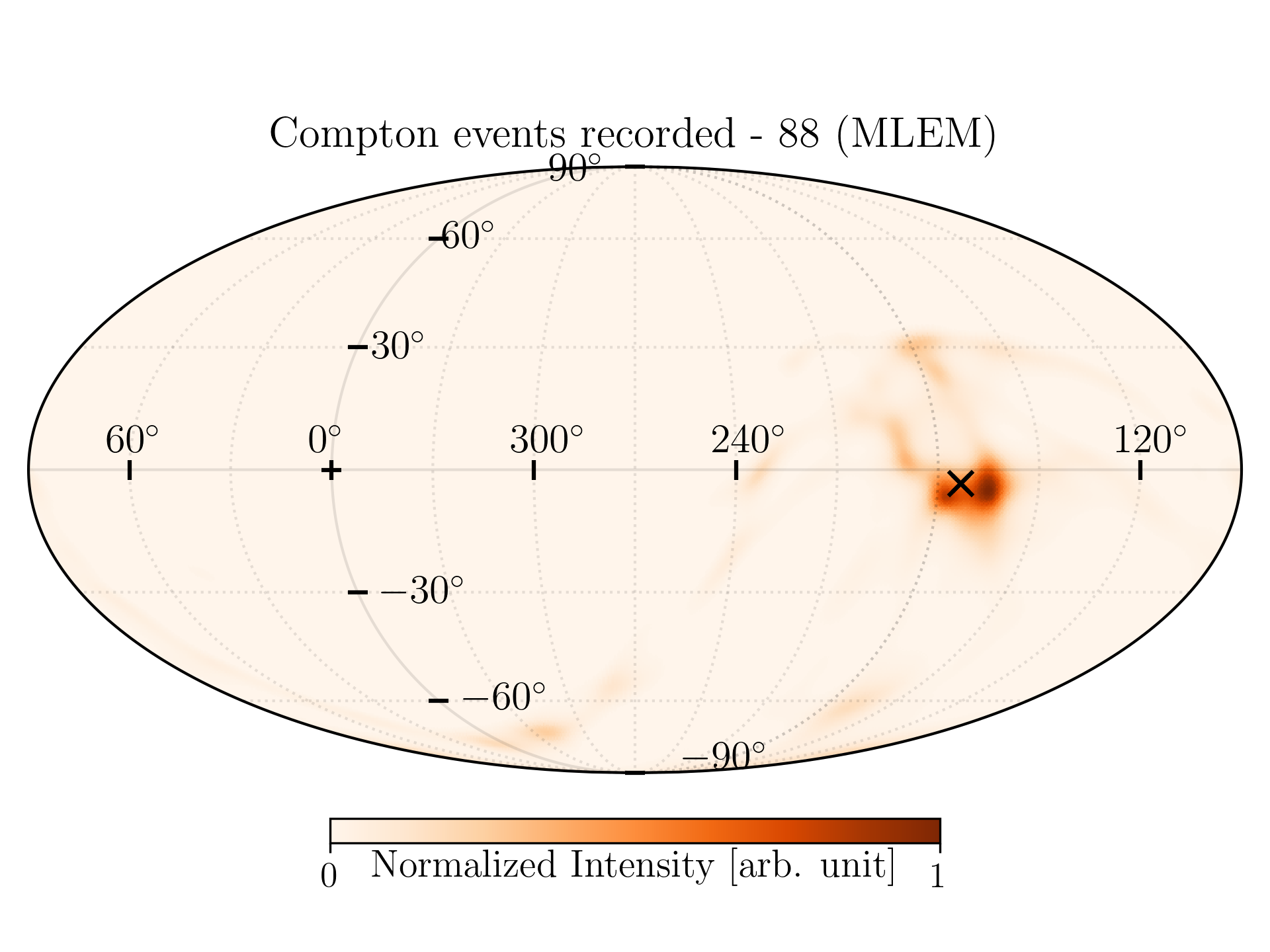} \hfill
    \includegraphics[width=0.45\textwidth]{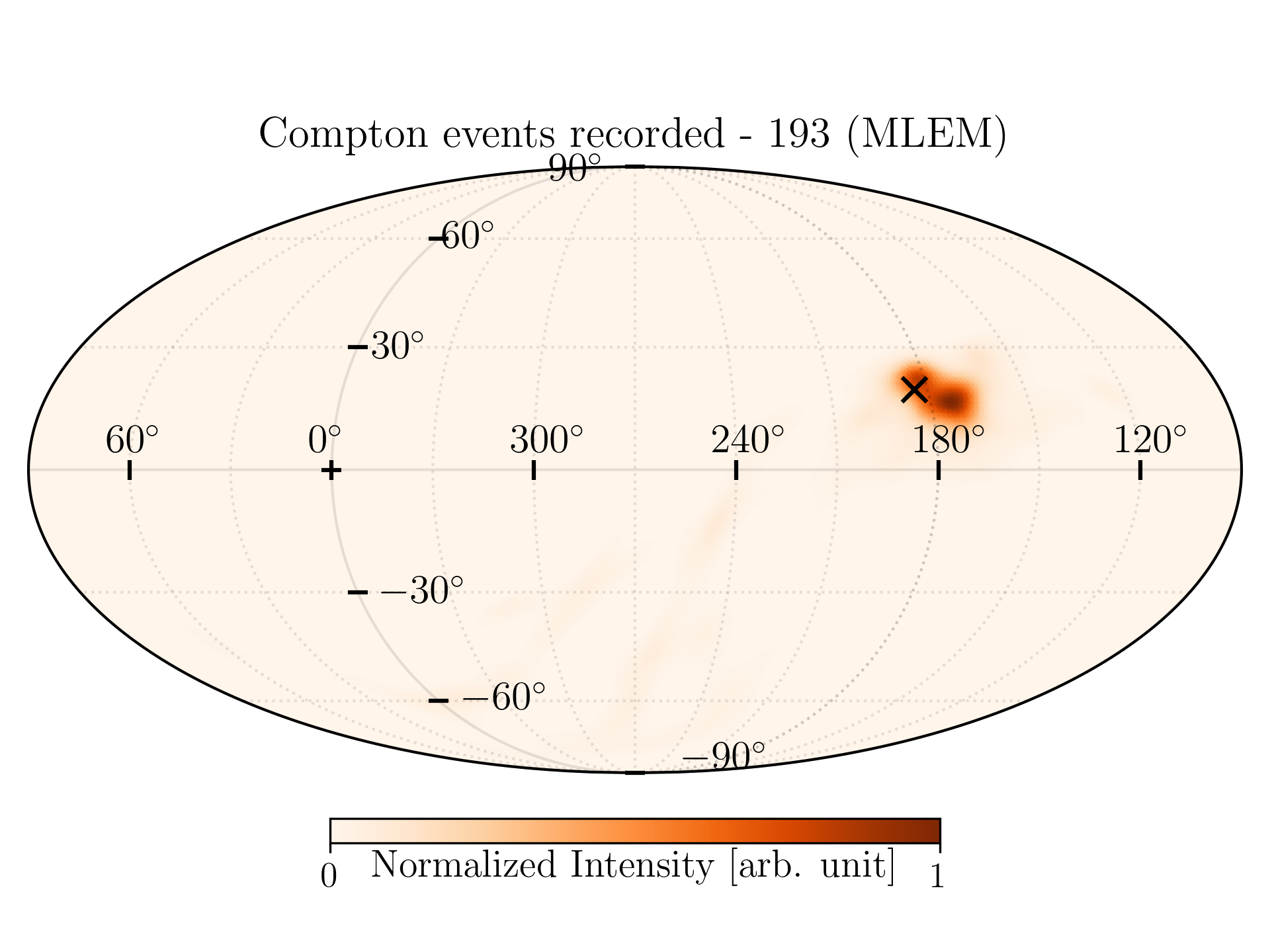}\\
    \includegraphics[width=0.45\textwidth]{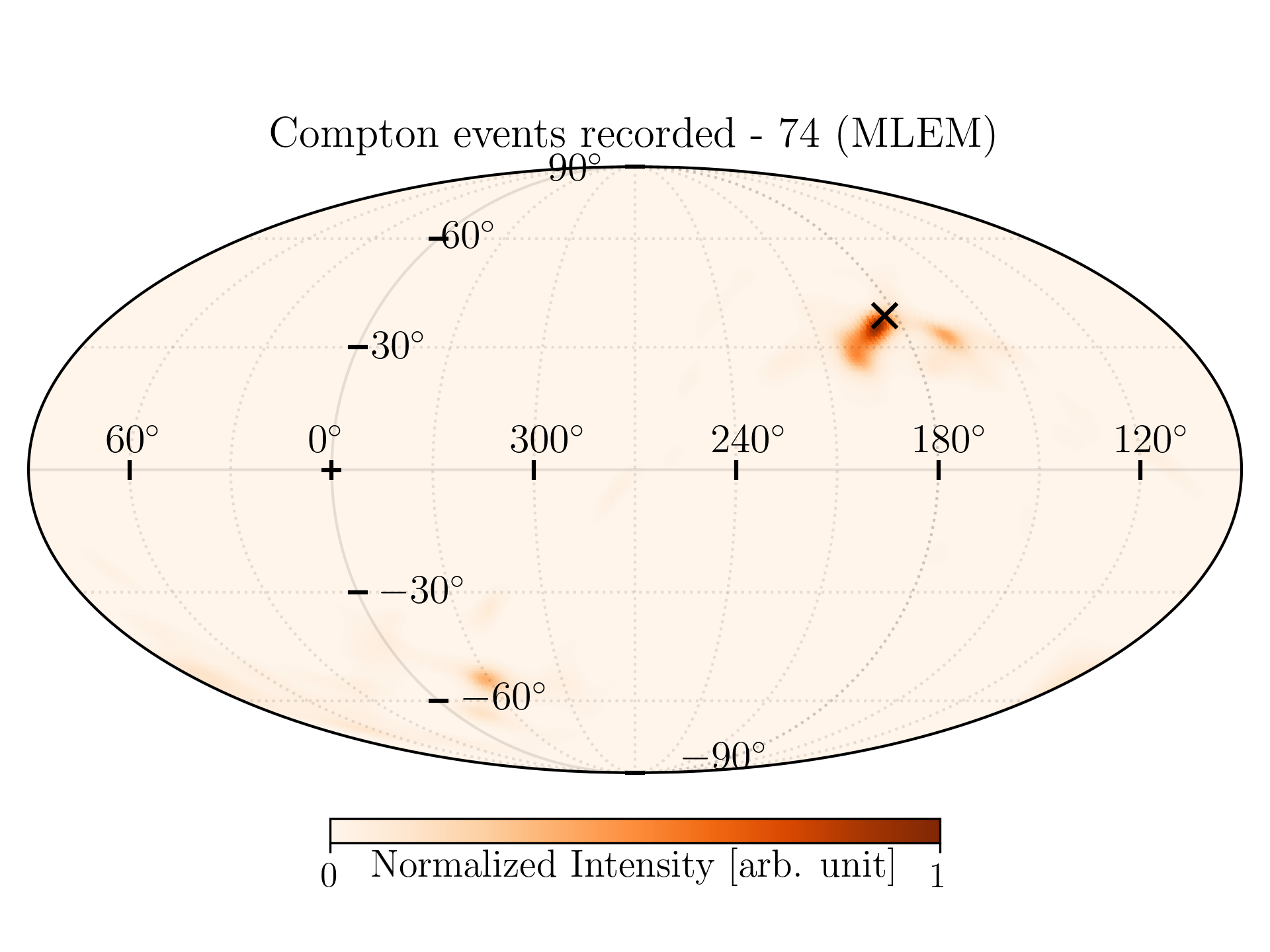} \hfill
    \includegraphics[width=0.45\textwidth]{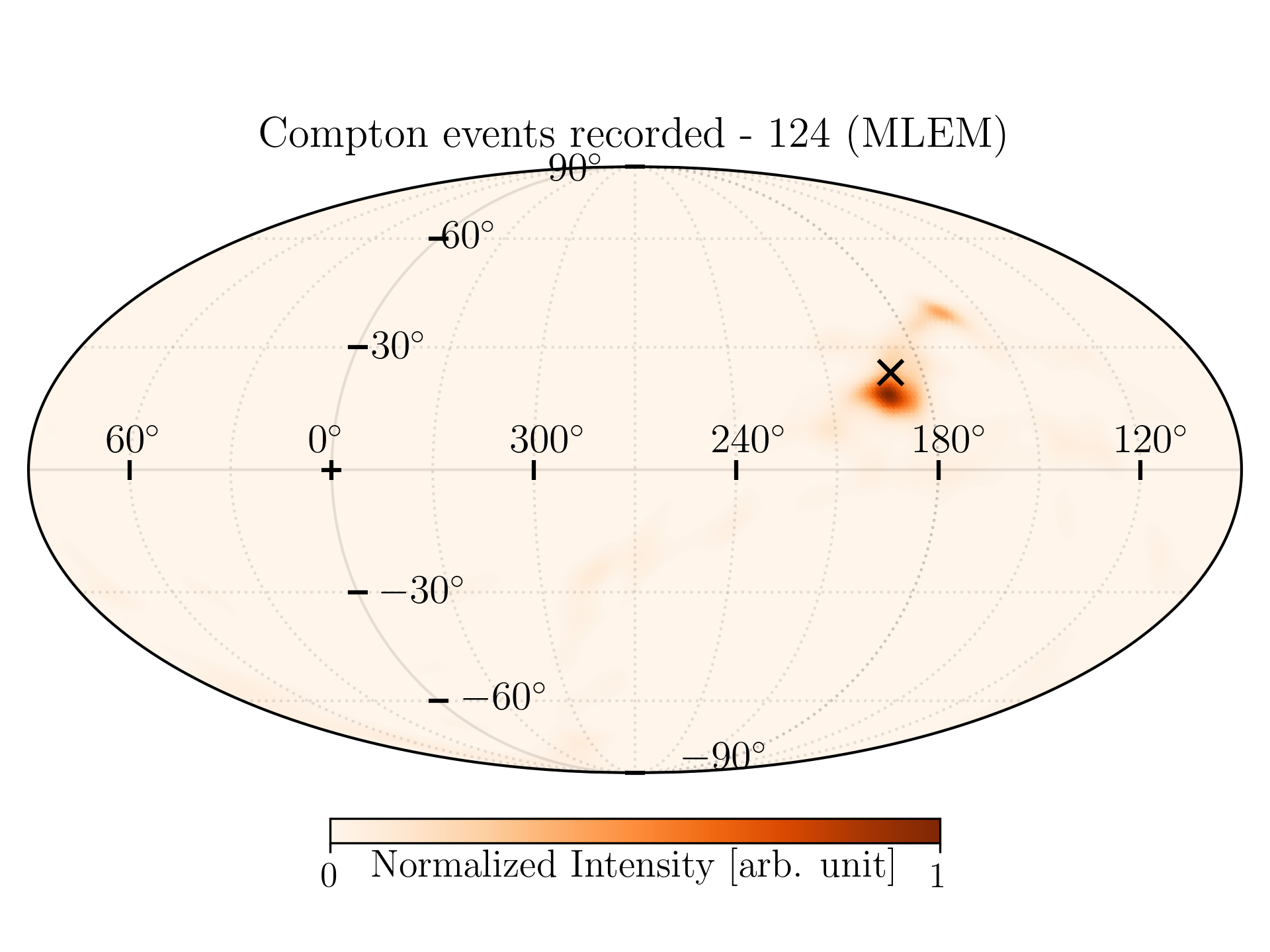}
     \caption{\update{All-sky maps displaying the 8th iteration of the MLEM algorithm. Localization has greatly improved as compared to the back-projection method (Figure \ref{fig:skymaps}) and has restricted itself in the inside of the field of view of the source location. The actual source location (marked by a cross) is consistent with the localization.
    \textit{Upper left:} Case 1 ($\theta = -3\degr, \phi=173\degr$). \textit{Upper right:} Case 2 ($\theta = 20\degr, \phi=184\degr$). \textit{Lower Left:}  Case 3 ($\theta = 38\degr, \phi=184\degr$). \textit{Lower Right:} Case 4 ($\theta = 24\degr, \phi=190\degr$).}}
    \label{fig:skymaps_MLEM}
\end{figure*}

\update{Figure \ref{fig:skymaps_MLEM} shows the results of eight iterations of the MLEM algorithm. In all cases, the results are consistent and the localization is much improved with little variation outside in the field of view outside of the source location.}

\subsection{Limitations}\label{subsec:limitations}
Our setup can satisfactorily recover the position of a \element{133}{Ba} source. However, there are certain limitations which impact the overall performance of the system, which we discuss below.

\subsubsection{Limited absorber coverage}\label{sec:lim_coverage}~\\
Source photons incident on Det~V can be scattered in any direction. Only a fraction of these photons will hit Det~H, while those travelling in other directions will not be detected. To mitigate the problem, one can add more detectors to the system.

The addition of more detectors can also allow a more diverse range of scatterer--absorber vectors ($\overline{g_1}$, Figure~\ref{fig:compton_imaging_working}). A simple interpretation of this is an increased spread in the centers of the event rings, thereby improving source localization \citep{zoglauerthesis}.

\subsubsection{Geometrical Error}\label{sec:lim_geom}~\\
The individual pixels of our CZT detectors are $2.46 \times 2.46 \times 5~\mathrm{mm}^3$ in size. We cannot infer the exact position of the photon interaction in this volume. Coupled with the relatively small (tens of mm) distance between the scattering and absorbing pixels, this leads to a few-degree uncertainty in the scattering angle, \update{with a nominal value of $\sim 2.5\degr$ for pixels near the detector center}.

This error can be reduced within the current setup by increasing the separation between the detectors. But in that scenario the absorber subtends a smaller solid angle from the scatterer compounding the problem in \S\ref{sec:lim_coverage}.

\subsubsection{Chance Coincidences}\label{sec:chance_coin}~\\
As discussed in \S\ref{sec:compsel}, the energies most of the temporally coincident photon pairs do not add up to any line emitted by the source. Some of these may arise from Compton scattering of lower energy lines of \element{133}{Ba}, but most are likely to be chance coincident photons rather than being true Compton event pairs. In our four tests, only about 4--7\%\ of the temporally coincident photons came from the 356~keV line. A consequence of this is the need to select photons of a known source energy, which might not always be feasible.

There are two methods to mitigate this: decreasing the background rate, and decreasing the chance coincidence probability. The spectra for individual detectors (Figure~\ref{fig:e0_vs_e1}) are consistent with being dominated by background noise near the LLD. This can be mitigated by cooling the detectors to a lower temperature. The detectors were operated with a 10~MHz clock speed in our tests, but they support higher operations speeds upto 30~MHz. Tripling the readout rate will decrease the chance coincidence probability by a factor of three, further improving the fraction of real Compton events. These two are the most important steps in converting our setup into a broad application Compton Imager where we can relax the source energy assumption.

\subsubsection{Energy Resolution}~\\
Our detectors operating at room temperature have a resolution of 12\%\ at 59.6~keV, and about 3.2\%\ at 356~keV. Poorer energy resolution increases the width of the event rings through Equation~\ref{eq:delta_psi_E}, thereby limiting angular resolution. For the 356~keV line in our configuration, this value is about 2\degr, comparable to the geometrical error $\Delta \psi_P$ discussed  \S\ref{sec:evt_circles} and \S\ref{sec:lim_geom}. This can be improved upon by either cooling the detectors, or by switching to other detectors. \update{The effect of cooling on the angular resolution will be limited: based on typical temperature dependence of energy resolution for solid state detectors, cooling the detectors by 10--20\degr C will improve the resolution by $<1\%$, and the corresponding contribution to the ring width will remain $>1.5\%$. On the other hand, switching detectors may help: }for instance, {\em NuSTAR} detectors have a resolution of 1\%\ at 59.6~keV \citep{2011SPIE.8145E..07K} and would give a much lower error. However, they are thinner (2~mm) and would not be sensitive to the higher energy photons used in this experiment.

\subsubsection{Source Position}~\\
In our experiment, the source was placed at distances of 22--28~cm from the 4~cm--wide scatterer. This causes a divergence of $\sim 10\degr$ in the source beam. However, our reconstruction algorithms assume incoming photons are parallel to each other. This ``near field source error'' is likely the largest contributor to the high ARM seen in Figure~\ref{fig:arm_plots}. Taking the source further away decreases the source count rate, and was not feasible within the practical constraints of a laboratory setup. \update{With a distant source, this beam divergence term will go to zero, and the ARM should improve by up to about ten degrees subtracted in quadrature.}

\subsubsection{Doppler broadening}~\\
Equation \ref{eq:scat_angle} was derived based on the assumption that the target electron is at rest. However, in the detector the electrons are bound to a nucleus and have finite momentum. This introduces another uncertainty in the calculated energies which fundamentally limits the achievable angular resolution. This effect, called as Doppler broadening, is highly dependent on detector's atomic number, \updatetwo{Compton scatter angle} and incident photon energy. 
\updatetwo{\cite{zoglauerthesis} states that the FWHM of the ARM due to Doppler broadening in CZT detectors follows a power-law dependence on the incident photon energy, and values for several energies are tabulated in their Table~2.1. Interpolating from those values, for our setup we get an angular resolution limit of $\sim 2.1^{\circ}$ at 356~keV.}

%
\section{Conclusion and Future Work}\label{sec:conc}

We have successfully demonstrated a laboratory setup of a Compton imager in the hard X-ray energy range. 

We developed a new system based on simple custom PCBs and an off--shelf PYNQ board to operate the detectors and read out data into an FPGA or a PC. With this setup, we achieved detector performance on par with those of \asat/CZTI. \update{There are two key updates as compared to CZTI. First, the $20~\mu$s timing resolution of CZTI would have given us a high chance coincidence background, hence we operate the detectors with a higher time resolution of $7.5~\mu$s. Second, detectors in CZTI were polled serially. Here, we upgraded the setup and have successfully demonstrated the ability to concurrently read out two detectors, paving the way for concurrent readouts of a larger number of detectors in future experiments.}

We tested the imager by irradiating it with \element{133}{Ba} from four different positions. In each case, we could successfully recover the source position from the recorded event data. In our tests, we obtain a direction--dependent angular resolution in the range of $\sim 16\degr$ to $\sim 30\degr$. The resolution is primarily limited by the ``near field'' location of the source, which is not factored into the reconstruction algorithms. The overall angular resolution and reconstruction performance can also be improved by using more advanced position reconstruction algorithms.

\update{The current setup is simplistic and has some limitations, discussed in \S\ref{subsec:limitations}, which limit the system performance. We have outlined possible technical improvements on several aspects, which were unfortunately not feasible within this setup. In the future, it will be worth operating the setup at a cooler temperature, and at a higher clock speed. The system can also be expanded to include more detectors in tandem, increasing the absorber coverage as well as increasing the overall effective area of the imager.
}

This successful technology development paves the way for the development of future complex CZT detector systems. The experience gained and lessons learnt from this experiment have aided the development of the Medium Energy Detector Package for the proposed \emph{Daksha} space telescope \citep{2024ExA....57...24B}.


\section*{Acknowledgements}
This work was supported by the joint IIT Bombay -- Indian Space Research Organisation (ISRO) Space Technology Cell (STC). We thank the staff of IIT Bombay and the Tata Institute of Fundamental Research (TIFR) for their help in setting up and running the experiments. We thank Ayush Nema for useful discussions. \update{We thank the anonymous referee for useful suggestions.}









\bibliography{References}{}

\begin{thebibliography}{}
\expandafter\ifx\csname natexlab\endcsname\relax\def\natexlab#1{#1}\fi

\bibitem[{{Bhalerao} {$et~al$.}(2017){Bhalerao}, {Bhattacharya}, {Vibhute},
  {Pawar}, {Rao}, {Hingar}, {Khanna}, {Kutty}, {Malkar}, {Patil}, {Arora},
  {Sinha}, {Priya}, {Samuel}, {Sreekumar}, {Vinod}, {Mithun}, {Vadawale},
  {Vagshette}, {Navalgund}, {Sarma}, {Pandiyan}, {Seetha}, \&
  {Subbarao}}]{Bhalerao2017}
{Bhalerao}, V., {Bhattacharya}, D., {Vibhute}, A., {$et~al$.} 2017, Journal of
  Astrophysics and Astronomy, 38, 31

\bibitem[{{Bhalerao} {$et~al$.}(2024){Bhalerao}, {Vadawale}, {Tendulkar},
  {Bhattacharya}, {Rana}, {Adalja}, {Belatikar}, {Bhaganagare}, {Dewangan},
  {Ghodgaonkar}, {Kumar Goyal}, {Gunasekaran}, {P J}, {Koyande}, {Kulkarni},
  {Kutty}, {Ladiya}, {Mahapatra}, {Marla}, {Mate}, {Mithun}, {Mote}, {Narang},
  {Nema}, {Nimbalkar}, {Pai}, {Palit}, {Patel}, {Patel}, {Pradeep},
  {Ramachandran}, {Bharath Saiguhan}, {Saraogi}, {Sawant}, {Shanmugam},
  {Sharma}, {Shetye}, {Singh}, {Singh}, {Singhal}, {Sreekumar}, {Sridhar},
  {Srinivasan}, {Tallur}, {Tiwari}, {Lakshmi Vadladi}, {Vaishnava},
  {Vishwakarma}, \& {Waratkar}}]{2024ExA....57...24B}
{Bhalerao}, V., {Vadawale}, S., {Tendulkar}, S., {$et~al$.} 2024, Experimental
  Astronomy, 57, 24

\bibitem[{{Boggs} \& {Jean}(2000)}]{2000A&AS..145..311B}
{Boggs}, S.~E., \& {Jean}, P. 2000, \aaps, 145, 311

\bibitem[{Compton(1923)}]{Compton1923}
Compton, A.~H. 1923, Phys. Rev., 22, 409

\bibitem[{Frandes {$et~al$.}(2016)Frandes, Timar, \& Lungeanu}]{Frandes2016}
Frandes, M., Timar, B., \& Lungeanu, D. 2016, Current Medical Imaging, 12, 95

\bibitem[{{Kierans}(2018)}]{kieransthesis}
{Kierans}, C. 2018, PhD thesis, University of California, Berkeley

\bibitem[{{Kierans} {$et~al$.}(2022){Kierans}, {Takahashi}, \&
  {Kanbach}}]{kierans2022}
{Kierans}, C., {Takahashi}, T., \& {Kanbach}, G. 2022, in Handbook of X-ray and
  Gamma-ray Astrophysics (Springer Nature Singapore), 18

\bibitem[{Kinsey {$et~al$.}(1996)}]{kinsey1996nudat}
Kinsey, R.~R., {$et~al$.} 1996, The NUDAT/PCNUDAT Program for Nuclear Data,
  Paper submitted to the 9th International Symposium of Capture Gamma-Ray
  Spectroscopy and Related Topics, Budapest, Hungary, October 1996. Data
  extracted from the NUDAT database, version (Aug 2024)

\bibitem[{{Kitaguchi} {$et~al$.}(2011){Kitaguchi}, {Grefenstette}, {Harrison},
  {Miyasaka}, {Bhalerao}, {Cook}, {Mao}, {Rana}, {Boggs}, \&
  {Zoglauer}}]{2011SPIE.8145E..07K}
{Kitaguchi}, T., {Grefenstette}, B.~W., {Harrison}, F.~A., {$et~al$.} 2011, in
  Society of Photo-Optical Instrumentation Engineers (SPIE) Conference Series,
  Vol. 8145, Society of Photo-Optical Instrumentation Engineers (SPIE)
  Conference Series, 814507

\bibitem[{{Knoll}(2000)}]{knoll}
{Knoll}, G.~F. 2000, {Radiation detection and measurement} (John Wiley and Sons
  (WIE))

\bibitem[{{Kotoch} {$et~al$.}(2011){Kotoch}, {Nandi}, {Debnath}, {Malkar},
  {Rao}, {Hingar}, {Madhav}, {Sreekumar}, \&
  {Chakrabarti}}]{2011ExA....29...27K}
{Kotoch}, T.~B., {Nandi}, A., {Debnath}, D., {$et~al$.} 2011, Experimental
  Astronomy, 29, 27

\bibitem[{Li {$et~al$.}(2024)Li, Cheng, Liu, Wang, Wen, Huang, \&
  Wu}]{s24030725}
Li, Z., Cheng, J., Liu, F., {$et~al$.} 2024, Sensors, 24, doi:10.3390/s24030725

\bibitem[{Nandi {$et~al$.}(2009)Nandi, Rao, Chakrabarti, Malkar, Sreekumar,
  Debnath, Hingar, Kotoch, Kotovk, \&
  Arkhangelskiy}]{nandi2009indianpayloadsrt2experiment}
Nandi, A., Rao, A.~R., Chakrabarti, S.~K., {$et~al$.} 2009, Indian Payloads
  (RT-2 Experiment) Onboard CORONAS-PHOTON Mission, arXiv:0912.4126

\bibitem[{Ramagond {$et~al$.}(2017)Ramagond, Yellampalli, \&
  Kanagasabapathi}]{ele8358511}
Ramagond, S., Yellampalli, S., \& Kanagasabapathi, C. 2017, in 2017
  International Conference On Smart Technologies For Smart Nation
  (SmartTechCon), 946--951

\bibitem[{{Sankarasubramanian} {$et~al$.}(2017){Sankarasubramanian},
  {Sudhakar}, {Nandi}, {Ramadevi}, {Adoni}, {Kushwaha}, {Agarwal}, {Dey},
  {Joshi}, {Singh}, {Girish}, {Tomar}, {Majhi}, {Olekar}, {Bug}, {Pala},
  {Thakur}, {Badagandi}, {Ravishankar}, {Garg}, {Sitaramamurthy}, {Sridhara},
  {Umapathy}, {Gupta}, {Agrawal}, \& {Yougandar}}]{2017CSci..113..625S}
{Sankarasubramanian}, K., {Sudhakar}, M., {Nandi}, A., {$et~al$.} 2017, Current
  Science, 113, 625

\bibitem[{{Schoenfelder} {$et~al$.}(1993){Schoenfelder}, {Aarts}, {Bennett},
  {de Boer}, {Clear}, {Collmar}, {Connors}, {Deerenberg}, {Diehl}, {von
  Dordrecht}, {den Herder}, {Hermsen}, {Kippen}, {Kuiper}, {Lichti},
  {Lockwood}, {Macri}, {McConnell}, {Morris}, {Much}, {Ryan}, {Simpson},
  {Snelling}, {Stacy}, {Steinle}, {Strong}, {Swanenburg}, {Taylor}, {de Vries},
  \& {Winkler}}]{schonfelder1993}
{Schoenfelder}, V., {Aarts}, H., {Bennett}, K., {$et~al$.} 1993, \apjs, 86, 657

\bibitem[{{Tomsick} {$et~al$.}(2019){Tomsick}, {Zoglauer}, {Sleator}, {Lazar},
  {Beechert}, {Boggs}, {Roberts}, {Siegert}, {Lowell}, {Wulf}, {Grove},
  {Phlips}, {Brandt}, {Smale}, {Kierans}, {Burns}, {Hartmann}, {Leising},
  {Ajello}, {Fryer}, {Amman}, {Chang}, {Jean}, \& {von
  Ballmoos}}]{tomsick2019cosi}
{Tomsick}, J., {Zoglauer}, A., {Sleator}, C., {$et~al$.} 2019, in Bulletin of
  the American Astronomical Society, Vol.~51, 98

\bibitem[{{Tomsick} {$et~al$.}(2024){Tomsick}, {Boggs}, {Zoglauer}, {Hartmann},
  {Ajello}, {Burns}, {Fryer}, {Karwin}, {Kierans}, {Lowell}, {Malzac},
  {Roberts}, {Saint-Hilaire}, {Shih}, {Siegert}, {Sleator}, {Takahashi},
  {Tavecchio}, {Wulf}, {Beechert}, {Gulick}, {Joens}, {Lazar}, {Neights},
  {Martinez Oliveros}, {Matsumoto}, {Melia}, {Yoneda}, {Amman}, {Bal}, {von
  Ballmoos}, {Bates}, {B{\"o}ttcher}, {Bulgarelli}, {Cavazzuti}, {Chang},
  {Chen}, {Chu}, {Ciabattoni}, {Costamante}, {Dreyer}, {Fioretti}, {Fenu},
  {Gallego}, {Ghirlanda}, {Grove}, {Huang}, {Jean}, {Khatiya},
  {Kn{\"o}dlseder}, {Kraus}, {Leising}, {Lewis}, {Lommler}, {Marcotulli},
  {Martinez Castellanos}, {Mittal}, {Negro}, {Al Nussirat}, {Nakazawa},
  {Oberlack}, {Palmore}, {Panebianco}, {Parmiggiani}, {Pike}, {Rogers},
  {Schutte}, {Sheng}, {Smale}, {Smith}, {Trigg}, {Venters}, {Watanabe}, \&
  {Zhang}}]{2024icrc.confE.745T}
{Tomsick}, J., {Boggs}, S., {Zoglauer}, A., {$et~al$.} 2024, in 38th
  International Cosmic Ray Conference, 745

\bibitem[{{von Ballmoos} {$et~al$.}(1989){von Ballmoos}, {Diehl}, \&
  {Schoenfelder}}]{1989A&A...221..396V}
{von Ballmoos}, P., {Diehl}, R., \& {Schoenfelder}, V. 1989, \aap, 221, 396

\bibitem[{Wilderman {$et~al$.}(1998)Wilderman, Clinthorne, Fessler, \&
  Rogers}]{wilderman1998list}
Wilderman, S.~J., Clinthorne, N.~H., Fessler, J.~A., \& Rogers, W.~L. 1998, in
  1998 IEEE Nuclear Science Symposium Conference Record. 1998 IEEE Nuclear
  Science Symposium and Medical Imaging Conference (Cat. No. 98CH36255),
  Vol.~3, IEEE, 1716--1720

\bibitem[{Yabu {$et~al$.}(2021)Yabu, Yoneda, Orita, Takeda, Caradonna,
  Takahashi, Watanabe, \& Moriyama}]{yabu2021tomographic}
Yabu, G., Yoneda, H., Orita, T., {$et~al$.} 2021, IEEE Transactions on
  Radiation and Plasma Medical Sciences, 6, 592

\bibitem[{{Zoglauer}(2005)}]{zoglauerthesis}
{Zoglauer}, A.~C. 2005, PhD thesis, Munich University of Technology, Germany

\end{thebibliography}

\end{document}